\documentclass[aps,prb]{revtex4}

\usepackage{epsfig,pstricks,dina4}

\usepackage{amsmath,amssymb}

\usepackage{graphicx}
\usepackage{epsfig,pstricks,amsmath,amssymb}

\def\RR{{\rm
         \vrule width.04em height1.58ex depth-.0ex
         \kern-.04em R}}
\def\i{{\rm i}\,}

\def\down{\downarrow}
\newcommand{\bigfrac}[2]{\mbox {${\displaystyle \frac{ #1 }{ #2 }}$}}

\newcommand{\bra}[1]{\left\langle #1 \right |}
\newcommand{\ket}[1]{\left | #1 \right\rangle}

\newcommand{\expect}[1]{\left\langle #1 \right\rangle}
\newcommand{\beq}{\begin{equation}}
\newcommand{\beqa}{\begin{eqnarray}}
\newcommand{\eeq}{\end{equation}}
\newcommand{\eeqa}{\end{eqnarray}}
\newcommand{\nbeqa}{\begin{eqnarray*}}
\newcommand{\neeqa}{\end{eqnarray*}}

\def\I {{\rm 1} \hspace{-1.1mm} {\rm I} \hspace{0.5mm}}

\begin{document}

\title{Entanglement in One-Dimensional Spin Systems}

\author{Andreas Osterloh, Luigi Amico}
\affiliation{MATISS-INFM $\&$ Dipartimento di Metodologie Fisiche e
    Chimiche (DMFCI), viale A. Doria 6, 95125 Catania, ITALY}
\author{Francesco Plastina, Rosario Fazio}
\affiliation{NEST-INFM $\&$ Scuola Normale Superiore, I-56127 Pisa, ITALY}

\begin{abstract}
For the anisotropic XY model in transverse magnetic field, we analyze the
ground state and its concurrence-free point for generic anisotropy,
and the time evolution of initial Bell states created in a fully polarized
background and on the ground state. 
We find that the pairwise entanglement
propagates with a velocity proportional to the reduced interaction 
for all the four Bell states.
A transmutation from singlet-like to triplet-like states is observed during the propagation.
Characteristic for the anisotropic models is the instantaneous
creation of pairwise entanglement from a fully polarized state;  
furthermore, the propagation of pairwise entanglement is suppressed 
in favor of a creation of different types of entanglement. 
The ``entanglement wave'' evolving from a Bell state on the ground state
turns out to be very localized in space-time.
Our findings agree with a recently formulated conjecture on entanglement 
sharing; some results are interpreted in terms of this conjecture. 
\end{abstract}

\maketitle

\section{Introduction}
A quantum mechanical system possesses additional correlations that do 
not have a classical counterpart. This phenomenon, called 
entanglement~\cite{Bell87}, is probably one of the most
astonishing features of quantum mechanics. Understanding the 
nature of these non-local correlations has been a central issue in the 
discussion of the foundation of quantum mechanics. More recently,
with the burst of interest in quantum information processing, entanglement 
has been identified as an ingredient for the speed-up in quantum 
computation and quantum communication protocols~\cite{Nielsen00} as 
compared with their classical counterparts. Therefore it is of crucial 
importance to be able to generate, manipulate and detect entangled states.
Experimental efforts in this direction have been put forward using photons~\cite{Zeilinger}, 
cavity QED systems~\cite{Haroche}, ion traps~\cite{Wineland}, 
and coupled quantum dots~\cite{Bayer}.
Also encouraged by the advances in the
field of nanoscience, there has been a number of proposals to 
detect signatures of entanglement. 
Systems under current study are multiterminal mesoscopic 
devices~\cite{Loss,Choi} and Josephson junctions~\cite{Plastina}.
\\
Though in many-body systems correlated states naturally appear,
a research activity investigating 
entanglement in condensed matter systems emerged only recently.
Despite conceptual difficulties, e.g. in distilling the quantum part of 
the correlations while the interaction between the subsystems is on, 
many-body systems might become useful for the development of 
new computation schemes and/or communication protocols. 
As an example we mention the recent proposal by Bose~\cite{Bose01}
to use the spin dynamics in Heisenberg rings to transfer 
quantum states. It is conceiveable that along the same lines other 
quantum information tasks can be implemented as well.

An important motivation for us to study the interconnection between 
condensed matter and quantum information is to investigate whether it 
is possible to gain additional insight in condensed matter states 
from quantum information theory~\cite{Preskill00}.
The peculiar aspects of non-local correlations become particularly 
evident when many bodies behave collectively; a prominent example is
a system close to a quantum phase transition~\cite{Sachdev00},
where it was found that entanglement can be classified in the framework of 
scaling theory~\cite{Osterloh02,Osborne02,Vidal02,Korepin03}, but
also profound differences between non-local quantum and 
classical correlations have been highlighted.
In a very recent paper the problem of decoherence in a 
near-critical one-dimensional system  was addressed~\cite{Khveshchenko03}.
The study of entanglement has not been devoted only to spin systems,
but also to the BCS model~\cite{Zanardi02,Delgado02}, 
quantum Hall~\cite{Zeng02,Balachandran97} and Boson systems~\cite{Hines02}.

In this paper, we examine the evolution of a local excitation bearing entanglement.
There are a number of questions that can be addressed in this way.
In particular we would like to see if there is a well defined velocity 
for the transport of entanglement and how it is related to the  
propagation of the elementary excitations of the spin system. 
Furthermore: is there a parameter regime which
favors entanglement transport even over larger distances; 
what are the time scales for the damping of entanglement created initially. 
Another important question to analyze is how the transport of entanglement
is influenced by the interference with other entangled states.
Finally, we try to discriminate pairwise from other types of entanglement.
This problem was quantified by Coffman, Kundu, and Wootters (CKW)
in terms of a conjecture on a measure for residual entanglement~\cite{Coffman00}. 
 
The paper is organized as follows: In the next 
section we will introduce the model Hamiltonian, its spectrum and 
correlation functions needed in the subsequent sections.
In section \ref{entgl} we present the applied entanglement measures. 
Section \ref{GS} is revisiting the ground state entanglement and
the entanglement dynamics for the isotropic and anisotropic model is 
presented in section \ref{dyn}. 
The final section is devoted to conclusions.

\section{The model}
The system under consideration is a spin-1/2 ferromagnetic chain with
an exchange coupling $\lambda$ in a transverse
magnetic field of strength $h$.
The Hamiltonian is $H=h H_s$ with the dimensionless Hamilton operator
$H_s$ being
\begin{equation}
H_s=-\lambda \sum_{i=1}^N (1+\gamma)S^x_i S^x_{i+1}+
(1-\gamma)S^y_i S^y_{i+1} - \sum_{i=1}^N S^z_i
\label{model}
\end{equation}
where $S^a$ are the spin-$1/2$ matrices ($a=x,y,z$) and $N$ is
the number of sites. We assume periodic boundary conditions.
The anisotropy parameter $\gamma$ connects the quantum Ising model
for $\gamma =1$ with the isotropic XY model for $\gamma = 0$.
In the interval  $0<\gamma\le 1$ the model belongs to the Ising
universality class and for $ N =\infty$ it undergoes a quantum phase
transition at the critical value $\lambda_c=1$. The order parameter is
the magnetization in $x$-direction, $\langle
S^x\rangle $, which is different from zero for $\lambda >1$ and
vanishes at and below the transition.
On the contrary the magnetization along the $z$-direction,
$\langle S^z\rangle $, is different from zero for any value of $\lambda$.
\\
This class of models can be diagonalized by means of the
Jordan-Wigner transformation~\cite{Lieb61,Pfeuty70,Mccoy70,Mccoy71}  that maps spins
to one dimensional spinless fermions with creation and annihilation operators
$c^\dagger_l$ and $c^{}_l$. It is convenient to use the operators
$A_l:= c_l^\dagger + c^{}_l$, $B_l:= c_l^\dagger - c^{}_l$,
which fulfill the anticommutation rules
$\{A_l, A_m\}=-\{B_l, B_m\}=2 \delta_{lm}$, 
$\{A_l, B_m\}=0$.
In terms of these operators the Jordan-Wigner transformation reads
$S_l^x=\frac{1}{2}A_l \prod_{s=1}^{l-1} A_s B_s$, 
$S_l^y=-\frac{i}{2}B_l \prod_{s=1}^{l-1} A_s B_s$, and
$S_l^z=-\frac{1}{2} A_l B_l $.
The Hamiltonian defined in Eq.(\ref{model}) is bilinear in the
fermionic degrees of freedom and is diagonalized
by means of the transformation
$\eta_k=\frac{1}{\sqrt{N}}\sum_l e^{ikl}[\alpha_k c^{}_l+i \beta_k c_l^\dagger ]$
with coefficients
$\alpha_k = \frac{\Lambda_k-(1+\lambda \cos k)}{\sqrt{2 [\Lambda_k^2
-(1+\lambda \cos k) \Lambda_k ]}}$ and
$\beta_k = \frac{\gamma \lambda \sin k}{\sqrt{2 [\Lambda_k^2
-(1+\lambda \cos k) \Lambda_k ]}}$.
The Hamiltonian then has the form
\begin{equation}
H=\sum_k \Lambda_k \eta_k^\dagger \eta_k
- \bigfrac{1}{2}\sum_k \Lambda_k\; ;\qquad 
   \Lambda_k= \sqrt{ \left (1+\lambda \cos k\right )^2 +
      \lambda^2 \gamma^2 \sin^2 k}\; .
\end{equation}

\subsection{Correlation functions}

As specified within the next section, one- and two site- entanglement
measures are obtained from the (one- and two-body) reduced density matrix
whose entries can be related to various spin correlation functions
\beq
M^{\alpha}_l(t) =\langle \psi | S_l^\alpha(t) |\psi  \rangle \; ;\qquad
g^{\alpha\beta}_{lm}(t) =\langle \psi | S_l^\alpha(t) S_m^\beta(t)| \psi  \rangle \; .
\label{correlators}
\eeq
These can be recast in the form of
Pfaffians~\cite{Mccoy70,Mccoy71,Caianello52}.
Correlators defined in Eq.(\ref{correlators}) have been calculated
for this class of models in the case of thermal
equilibrium~\cite{Lieb61,Pfeuty70,Mccoy70,Mccoy71}. In this
case, the expression for the correlators reduces to the
calculation of Toeplitz determinants. This is not the case here,
since the initial state $| \psi  \rangle$ is not an eigenstate of the Hamiltonian
Eq.(\ref{model}). 
The correlation functions needed here have been obtained in~\cite{AOXYdyn,AOXY}.
\\
As initial state we consider two sites in a Bell
state and all other spins being in the state $|\downarrow \rangle $ (or $|\uparrow \rangle $) 
\beq
\ket{\Psi_{i,j}^\varphi}  = 
\frac{1}{\sqrt{2}} \left (c_i^\dagger + e^{i\varphi} c_j^\dagger  \right ) 
\ket{\Downarrow}
\label{wavefunction}
\eeq
where the vacuum state is defined as $\ket{\Downarrow}=\ket{\down\dots\down}$.
This initial state explicitly breaks the translational invariance.
The correlators $\langle\Psi_{i,j}^\varphi| S_l^\alpha S_m^\beta
	|\Psi_{i,j}^\varphi\rangle$
can be expressed as a {\it sum of}\/ Pfaffians as descibed in~\cite{AOXYdyn,AOXY}
and allow, together with the magnetization,
to evaluate the two-body reduced density matrix.
Inaccessible by the above technique is the correlation 
$\bra{\pm} S_l^{x}\ket{\pm}$. 
However, if $\bra{\pm} S_l^{x,y}(t=0) \ket{\pm} =0$, then it will remain
zero during the subsequent evolution due to
the parity symmetry of the Hamiltonian.
We will consider exclusively this case in the present work.
In the case $\gamma=0$ the particle number conservation leads to a
considerable simplification since the vacuum is an eigenstate
and the correlation functions can be evaluated directly without resorting on
Pfaffians. For an open-ended isotropic chain dynamic correlators have been studied
employing the pfaffian expression~\cite{Stolze}.

\section{Measures of entanglement}\label{entgl}

On a qualitative basis entanglement is well understood. Both
for distinguishable particles (e.g. spins on a lattice) and
identical particles (e.g. free fermions/bosons).~\cite{Ghirardi02}
If a many-spin system is in a pure state, 
a rough measure for the entanglement between two subsystems is
the {\em mixedness} of the reduced density matrix of the subsystem.
We analyze the case when this subsystem is a single site and 
choose the one-tangle, which on site $j$ is given by
$$
\tau^{(1)}[\rho^{(1)}]:=4 {\rm det} \rho^{(1)}= \bigfrac{1}{4}-\expect{S^z_j}^2\; .
$$
Bipartite entanglement is encoded in the two-qubit reduced density matrix $\rho^{(2)}$,
obtained from the wave-function of the state after all the spins
except those at positions $i$ and $j$ have been traced out. The
resulting $\rho^{(2)}$ represents a mixed state of a bipartite system
for which a good deal of work has been devoted to quantify its
entanglement \cite{BennettBernstein96,Vedral97,BennettDiVincenzo96}. 
As a measure for mixed states of two qubits, we use the
concurrence~\cite{Wootters98} $C[\rho^{(2)}]$
\beq\label{def:concurrence}
C[\rho^{(2)}] := \max\{0,2\lambda_{max} - {\rm tr} \sqrt{R}\}\; ;\quad
R := \rho^{(2)} \sigma_y\otimes\sigma_y \rho^{(2)\,*}
\sigma_y\otimes\sigma_y  
\eeq 
where $\lambda_{max}$ is the largest eigenvalue of the matrix $\sqrt{R}$. 
For pure two-qubit states we have $\tau^{(1)}\equiv C^2=:\tau^{(2)}$. 
The concurrence, expressed in terms of spin correlation functions, is~\cite{AOXYdyn}
\beq\label{C-of-corrs}
C_{lm}=2\max\left\{
\sqrt{(g^{xx}_{lm}\mp g_{lm}^{yy})^2
    +(g_{lm}^{xy}\pm g_{lm}^{yx})^2}-
    \sqrt{(\bigfrac{1}{4}\mp g_{lm}^{zz})^2-\bigfrac{1}{4}(M^z_l\mp M^z_m)^2},0 \right\}
\eeq
In the isotropic case, additional constants of the motion simplify the
expression for the concurrence~\cite{AOXYdyn}.
\begin{figure}
\begin{center}
\includegraphics[width=.3\linewidth]{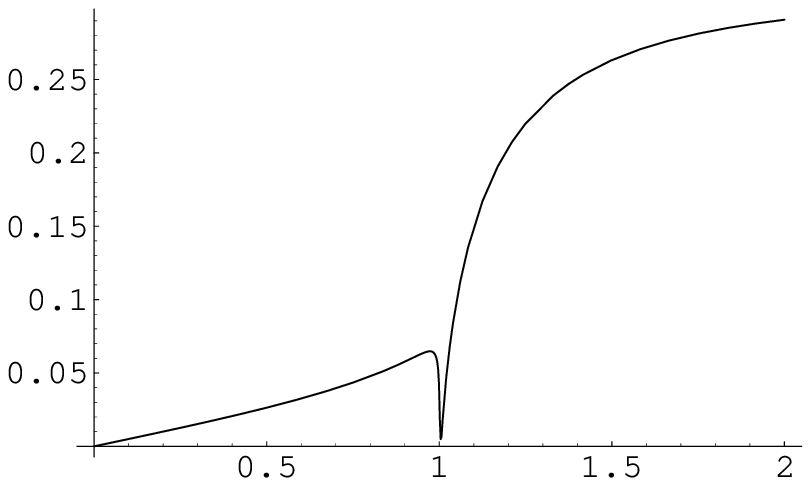}
\includegraphics[width=.3\linewidth]{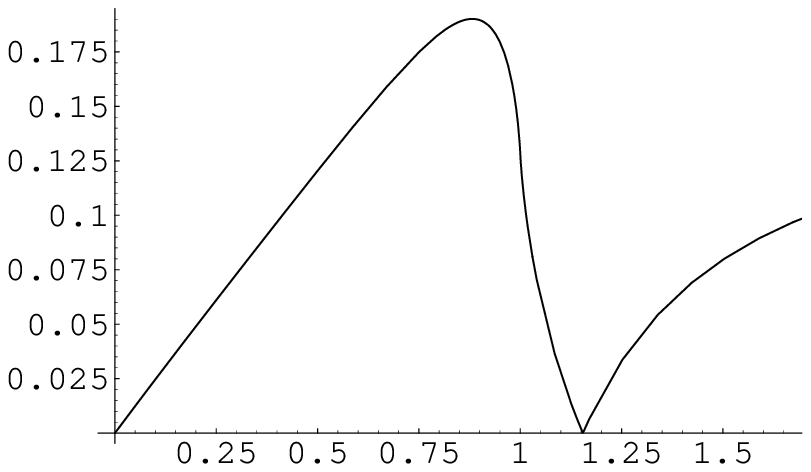}
\includegraphics[width=.3\linewidth]{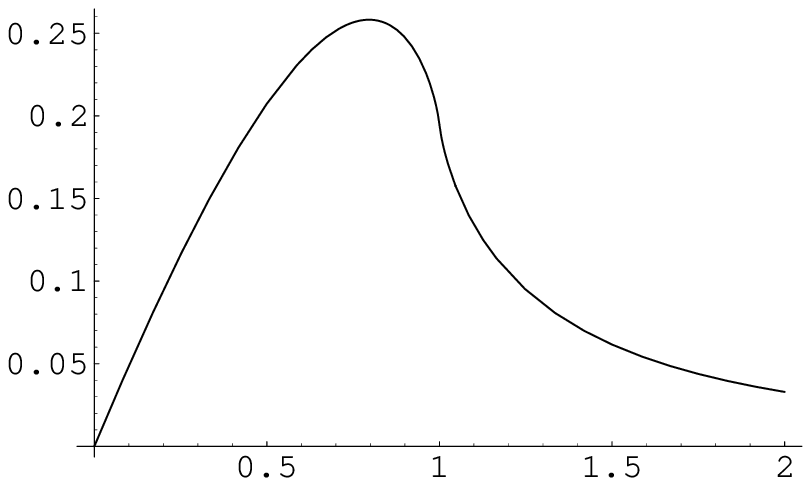}
\includegraphics[width=.28\linewidth,angle=-90]{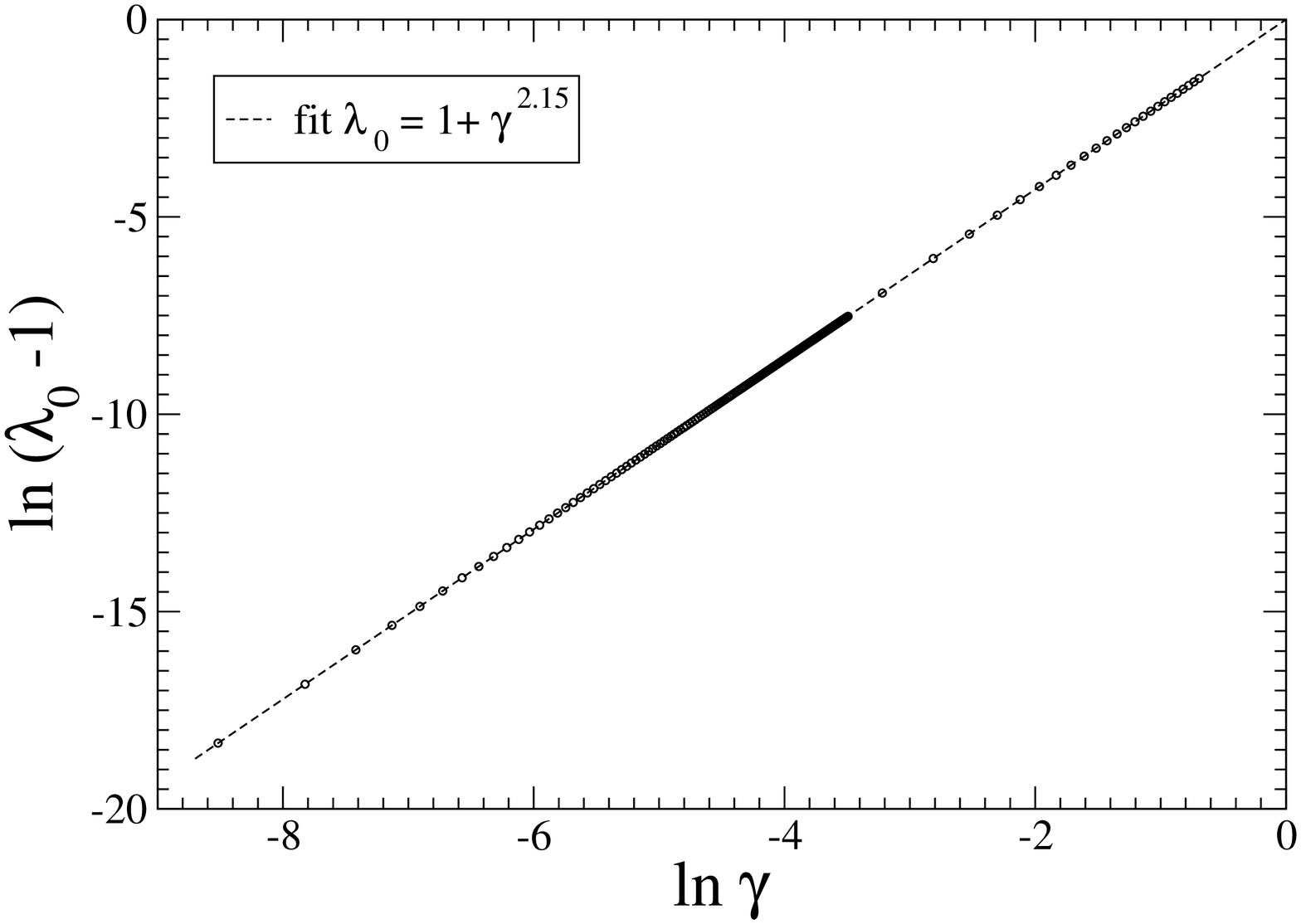}
\includegraphics[width=.28\linewidth,angle=-90]{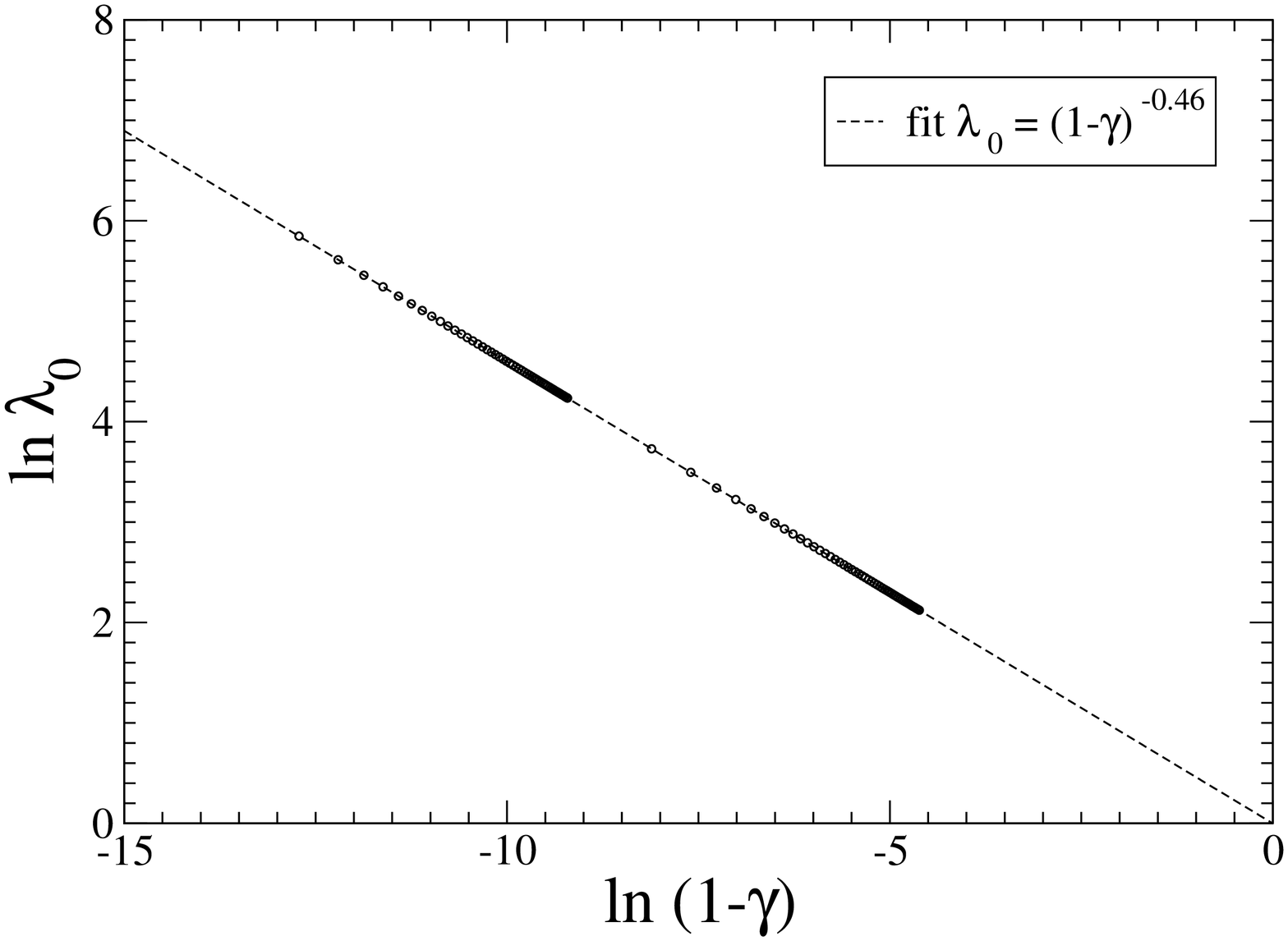}
\end{center}
\caption{{\em Top panel}: The nearest neighbor concurrence $C_1$ for the values of $\gamma=0.1$,
$0.5$, and $1$ (from left to right). At $\lambda=0.5$ and
$\lambda=1$, $C_1$ has roughly the same value: $0.0264$ and $0.0337$ for
$\gamma=0.1$, $0.1204$ and $0.1285$ for $\gamma=0.5$; $0.2074$ and $0.1946$
for $\gamma=1$, respectively. For the Ising model at $\lambda=0.9$ we have $C_1=0.24$.
These values remain as uniform background in the concurrence 
of the singlet-type perturbation of the ground state (see bottom panel of Fig.~\ref{C1-Vac}).
{\em Bottom panel}: The behavior of the separable point for generic $\gamma$ for
  {\em left:} $\gamma\rightarrow 0$ and 
  {\em right:} $\gamma\rightarrow 1$.}
\label{separable-point}
\end{figure}
One- and two- site entanglement do not furnish a complete characterization
of the entanglement present in spin chains. Following a conjecture put 
forward by Coffman, Kundu, and Wootters as follows
\beq\label{def:CKW}
\sum_{j\neq n} C^2_{n,j} \leq 4 \det \rho^{(1)}_n = \tau^{(1)}_n\; .
\eeq
it is interesting to study the difference of the quantities in Eq.~(\ref{def:CKW}),
interpreted as ``residual tangle'', i.e. entanglement not
stored in two-qubit entanglement.
This inequality was proved in Ref.~\cite{Coffman00}
for a three qubit system, giving rise to the definition
of the three-tangle as a measure of three-qubit entanglement.

\section{Ground state entanglement}\label{GS}

\begin{figure}
\begin{center}
\includegraphics[width=.3\linewidth,angle=-90]{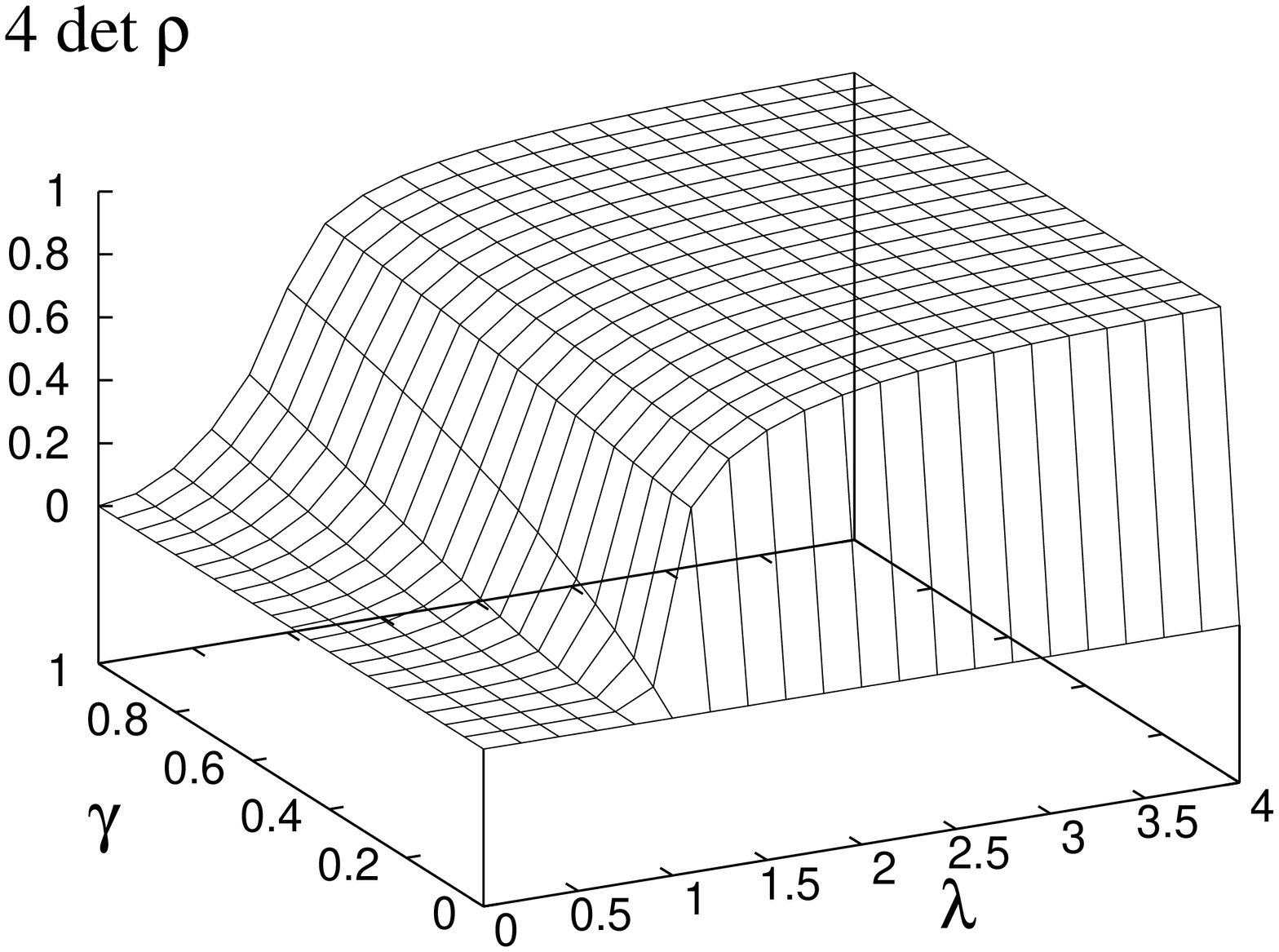}
\includegraphics[width=.27\linewidth,angle=-90]{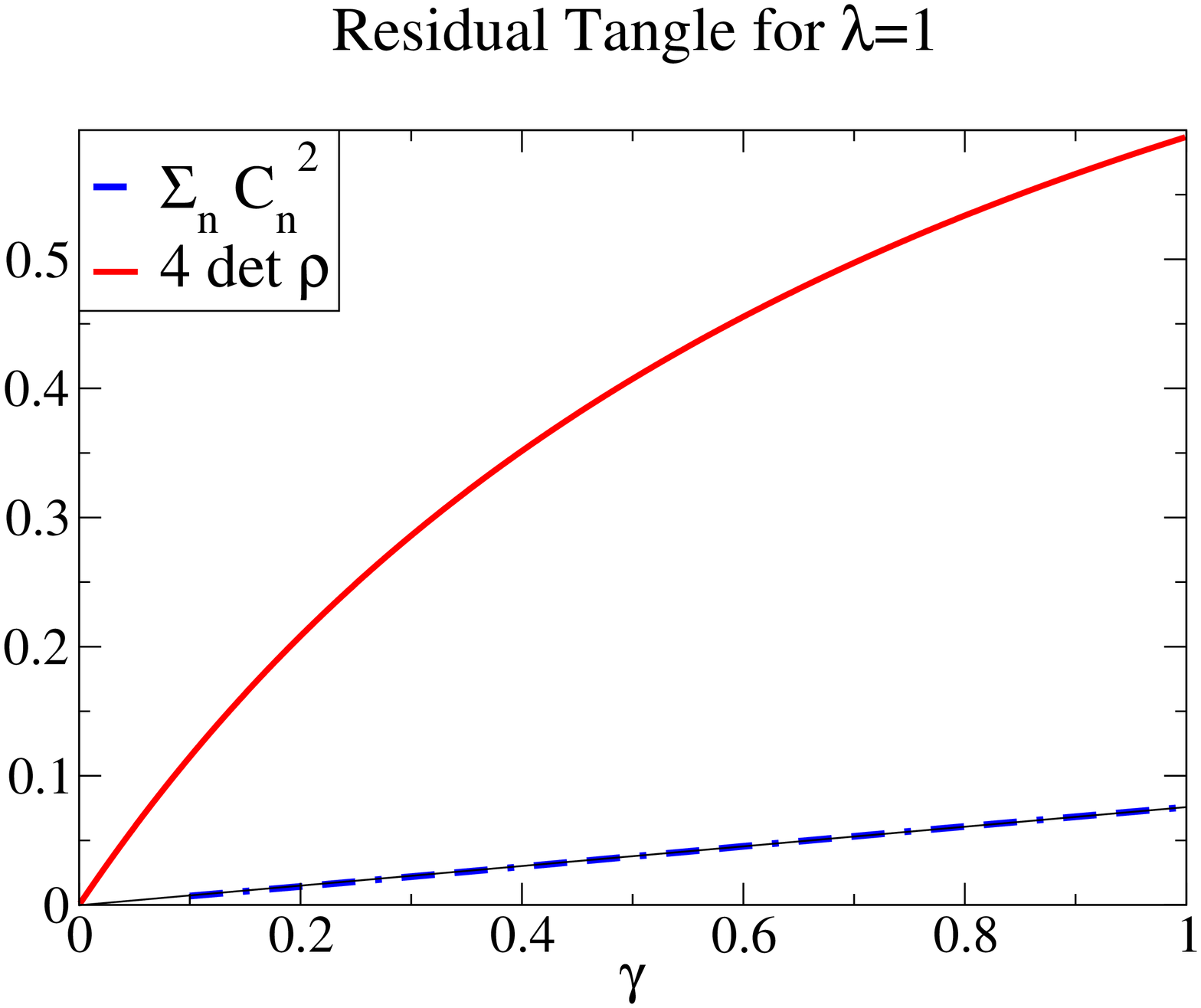}
\end{center}
\caption{The global tangle for the groundstate (left). 
For $0.1\leq\gamma\leq 1$
we could verify the CKW conjecture for critical coupling (right).
The one-tangle (thick line) is much larger than 
$\sum_{n=-\infty}^\infty C_n^2=2\sum_{n=1}^\infty C_n^2$ (thick dash-dotted line) 
suggesting that the major part of 
entanglement should be stored in higher than two-qubit entanglement. 
The thin line is a guide to the eye showing that the sum of the
two-tangles linearly tends to zero as $\gamma\rightarrow 0$.}
\label{4det-GS-g0upto1}
\end{figure}

This section focuses on the ground state entanglement.
In Fig.~\ref{separable-point}, upper panel, the nearest neighbor concurrence 
for the ground state is shown for different 
values of $\gamma$ as a function of the reduced coupling $\lambda$.
All exhibit a logarithmic divergence of the first 
derivative respect to $\lambda$ at the quantum critical point 
$\lambda_c=1$, and fit within finite size scaling theory.~\cite{Osterloh02}
This observation also applies to the first non-zero derivative
of the concurrence at larger distance.
The cusp, where $C_1$ vanishes is the point where the large eigenvalues
of $R$, Eq.(\ref{def:concurrence}), for the invariant sectors (due to the parity symmetry) 
$|M_z|=0$ and $|M_z|=1/2$ are equal.
This indicates a change in the type of Bell state responsible for the entanglement.
The value $\lambda_0$ where this degeneracy occurs (and the 2-qubit reduced
density matrix is separable) converges
from above to the critical coupling $\lambda_c=1$ for 
$\gamma\rightarrow 0$ as $\lambda_0=1+\gamma^{2.15}$ and diverges
as $\gamma\rightarrow 1$ as $\lambda_0=(1-\gamma)^{-0.46}$ (see Fig.~\ref{separable-point}).
In the ordered region, the thermic ground state is a mixed state
and as a consequence, the here shown concurrence is only an upper bound 
for the thermic ground state with spontaneously broken symmetry. 
However, it has been shown recently~\cite{Syljuasen} that if it is the 
triplet sector that
furnishes the largest eigenvalue of $R$, the concurrence is not
affected by the thermic mixing. Therefore, the symmetry breaking
does not affect the concurrence for $1\leq\lambda\leq \lambda_0$.
\\
The one-tangle for the ground state is shown on the left of figure
\ref{4det-GS-g0upto1}. 
At the critical point for $\gamma\in [0.1\, ,\, 1]$ we could demonstrate
that the sum of the two-tangles is much below the one-tangle.
This is seen in the right plot of figure \ref{4det-GS-g0upto1}.
If the CKW conjecture holds, this indicated that
far the most entanglement is stored in higher tangles (yet to be quantified).
This indication recently found further support coming from other indicators
for higher entanglement as 
the scaling of the von Neumann entropy of a compact block of spins
with the size of the block~\cite{Vidal02} and
the amount of localizable entanglement, quantified by the maximum 
two-point correlation function~\cite{Cirac03}.

\section{Dynamical evolution of an e-bit}\label{dyn}

The key to the time evolution of the original spin operators
is that of the spinless fermions
$c^{}_j(t)= \sum_l [\tilde{a}_{l-j}(t) c^{}_l - \tilde{b}_{l-j}(t)c^\dagger_l ]$,
where the new coefficients are
$\tilde{a}_{x}(t) = \frac{1}{\sqrt{N}}\sum_{k} \cos{k x}
    \left ( e^{i \Lambda_k t}-2i \beta^2_k \sin \Lambda_k t\right ) $ and\\
$\tilde{b}_{x}(t) = \frac{2\i}{\sqrt{N}}\sum_{k} \sin{k x}\,
    \alpha_k \beta_k \sin \Lambda_k t $.
In the limit $\gamma =0$ the previous expressions
simplify considerably. In this case the magnetization, i.e. the
$z$-component of the total spin $S^z=\sum_j S^z_j$, is a conserved
quantity. In terms of fermions this corresponds to the
conservation of the total number of particles, $N=\sum_j
n_j=\sum_j c^\dagger_j c^{}_j$. For $\gamma\rightarrow 0$ and
$|\lambda|\leq 1$ we find that $\alpha_k\rightarrow 0$ and
$\beta_k\rightarrow {\rm sign}\,k$. The energy spectrum is
$\Lambda_k=\left|1+\lambda \cos k\right|$ and the eigenstates are
plane waves 
$ c^{}_j(t) = \frac{1}{\sqrt{N}}\sum_{k}\sum_{l} \cos{k(l-j)} e^{-i \Lambda_k t} c^{}_l $
and
$\eta^\dagger_k=\frac{1}{\sqrt{N}}\sum_l e^{-ikl} c^{}_l$.

\subsection{$\gamma=0$: The isotropic model}
\label{section:gammazero}
In this section, we describe the dynamics of entanglement for
$\gamma=0$. The Hamiltonian of Eq.(\ref{model}) is then reduced to
the $XY$ model. Only in this case the $z$-component of the total spin, $S^z$, 
is conserved.
Consequently the Jordan-Wigner transformed fermionic Hamiltonian
becomes a tight binding model for each sector with fixed $S^z$.

\subsubsection{Propagation of states in the singlet sector}
\begin{figure}
\begin{center}
\includegraphics[width=.34\linewidth]{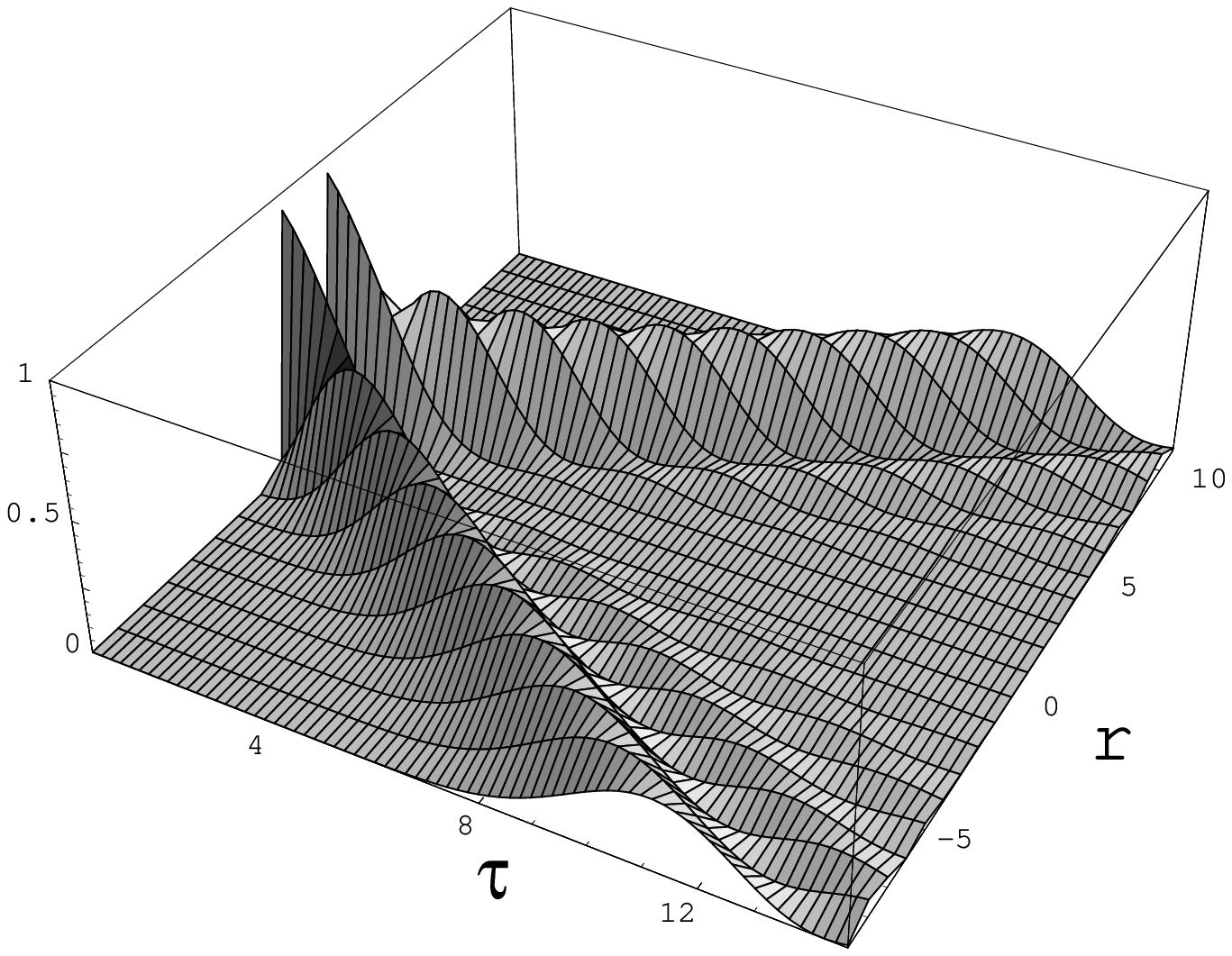}
\hspace*{17mm}
\includegraphics[width=.34\linewidth]{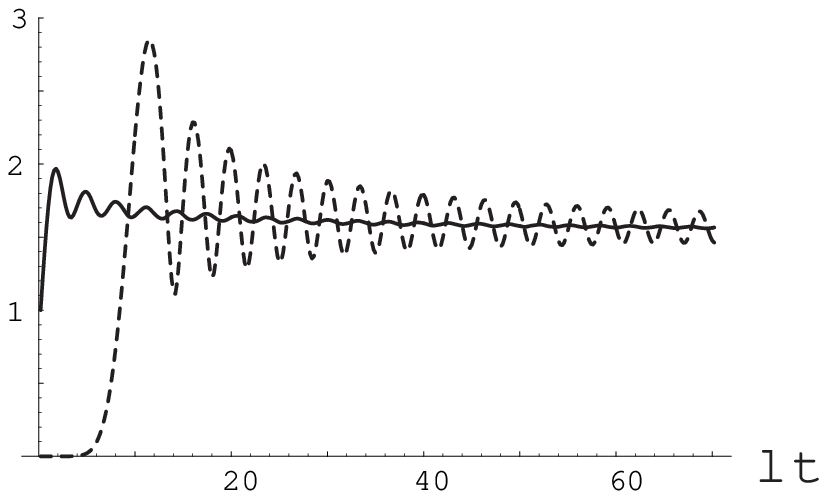}
\end{center}
\caption{{\em Left}: Concurrence between sites $n=-r, m=r$, symmetrically
displaced from their initial position $i=-1$ and $j=1$
($\varphi=\pi$). {\em Right}: Summed concurrences for an initially entangled site
($n=0$, full line) and an initially unentangled site ($n=10$,
dashed line). The initial state is a singlet ($\varphi=\pi$)
created on sites $i=0$ and $j=1$.} \label{treconc}
\end{figure}

We first consider the case of a chain initially prepared in a
maximally entangled singlet-like state $\ket{\Psi_{i,j}^{\varphi}}$ on sites 
$i$ and $j$ as defined in Eq. (\ref{wavefunction}). 
The state vector at later times is
$\ket{\Psi_{i,j}^\varphi} =\sum_l w_l(t) \, c_l^{\dagger} \, \ket{\Downarrow}$
with
$w_l(t) = \frac{1}{\sqrt{2N}} \sum_{k} [e^{\frac{2 \pi
i k }{N}(i-l)} + e^{i \varphi} e^{\frac{2 \pi i k }{N}(j-l)}] e^{i \Lambda _k t}$.~\cite{AOXYdyn}
For an infinite system, the coefficients become (up to a global phase)
$w_l(t) = \frac{1}{\sqrt{2}} \{
J_{i-l}(\lambda t) + e^{i \varphi} \,
 (-i)^{(j-i)} J_{j-l}(\lambda t) \}$,
where $J_n(x)$ is the Bessel function of order $n$.
For this initial singlet-like state
the concurrence of sites $n$ and $m$ is 
$C_{n,m}(t) = 2 | w_n(t) w_m^*(t) | $.
The time scale is set up by the interaction strength: the information
exchange or ``entanglement propagation" takes the time $t \sim d/\lambda$
for $d$ lattice spacings; i.e. the speed of propagation is $\lambda $.
The propagation of an EPR-like pair is demonstrated in Fig.\ref{treconc}, 
where the concurrence is shown between two
sites symmetrically displaced with respect to the initial
excitation ($C_{r,-r}(t)$ for $i=-j=1$) 
together with the sum of concurrence $C_{tot,n} = \sum_m C_{n,m}$
of the single site $n$. The latter tends to
the same stationary value (around 1.6) for arbitrary $n$, 
independently of the initial conditions\cite{AOXYdyn}. This means
that the initial state becomes homogeneously spread at long times
and so does the concurrence.

The same structure as for the concurrence is also found for the one-tangle.
This fits within the picture suggested by the CKW conjecture~\cite{Coffman00}.
For the isotropic model it is possible to analytically
check this conjecture. We obtain
$4 \det \rho^{(1)}_j=4 |w_j|^2 (1-|w_j|^2)$
and
$\sum_{n\neq j} C^2_{j,n}=4\sum_{n\neq j} |w_j w_n^*|^2
=  4 |w_j|^2 (1-|w_j|^2)$.
This has already been observed for these W-type states~\cite{Coffman00}.
Thus, the two quantities coincide, meaning that the entanglement
present in the system is restricted to the class of pairwise
entanglement, and no higher order entanglement is created. Therefore, 
the information in the one-tangle is already contained in the concurrence.

In order to gain information on what type of Bell state propagates
and on eventual state mutation, we study how similar is the mixed
state $\rho^{(2)}_{n,m}$ to the initial 
$\ket{\Psi^{\varphi}}$. 
This similarity is quantified by the fidelity
\begin{equation}
F_{n,m}(t) = \mbox{Tr} \left \{\rho^{(2)}_{n,m}(t) \,
\ket{\Psi^{\varphi}}\bra{\Psi^{\varphi}} \right \} = 
\frac{1}{2} \left | w_n(t) + e^{-i \varphi} w_m(t) \right |^2 \; .
\end{equation}
It was shown to display
in-phase oscillation with respect to one-tangle and concurrence~\cite{AOXYdyn}. 
Thus, when the entanglement wave arrives,
$\rho^{(2)}_{n,m}$ becomes more similar to the initially prepared
state. That means that the state itself is
propagating along the chain, taking with it its entanglement.
This propagation is far from being a perfect transmission, 
due to the entanglement sharing with many sites at a time.
\begin{figure}
\begin{center}
\includegraphics[width=.7\linewidth]{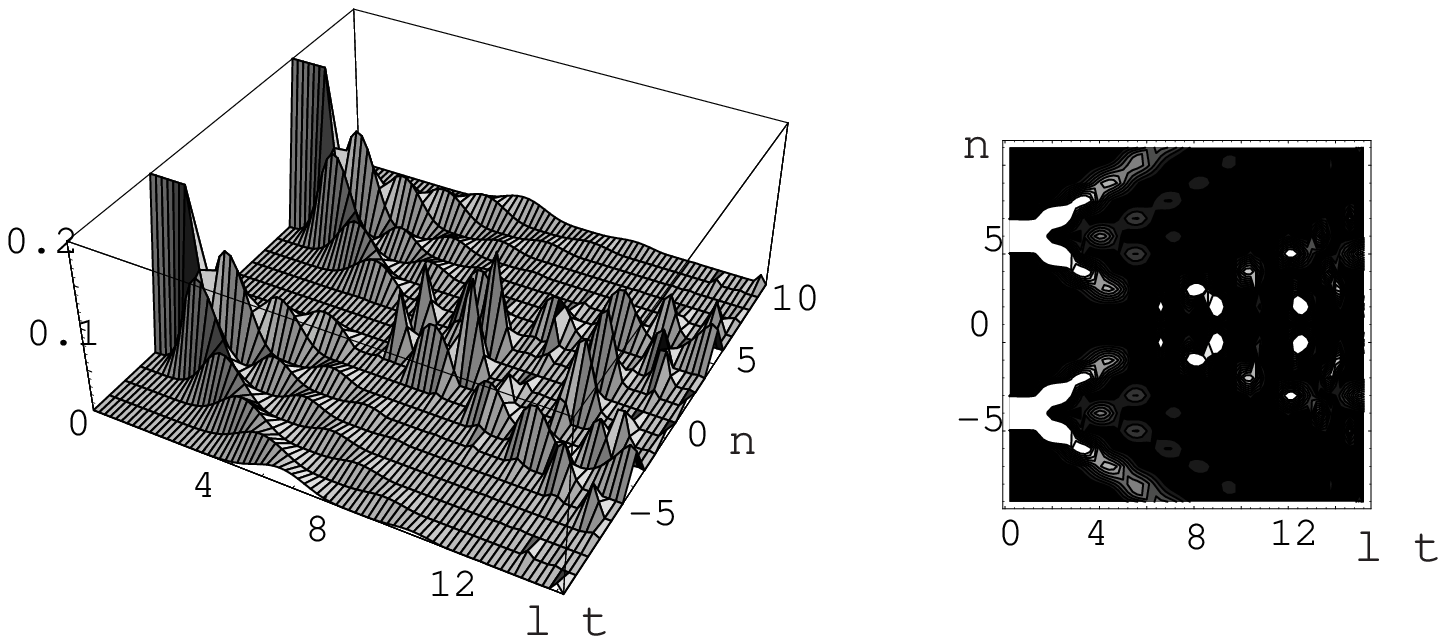}
\includegraphics[width=.7\linewidth]{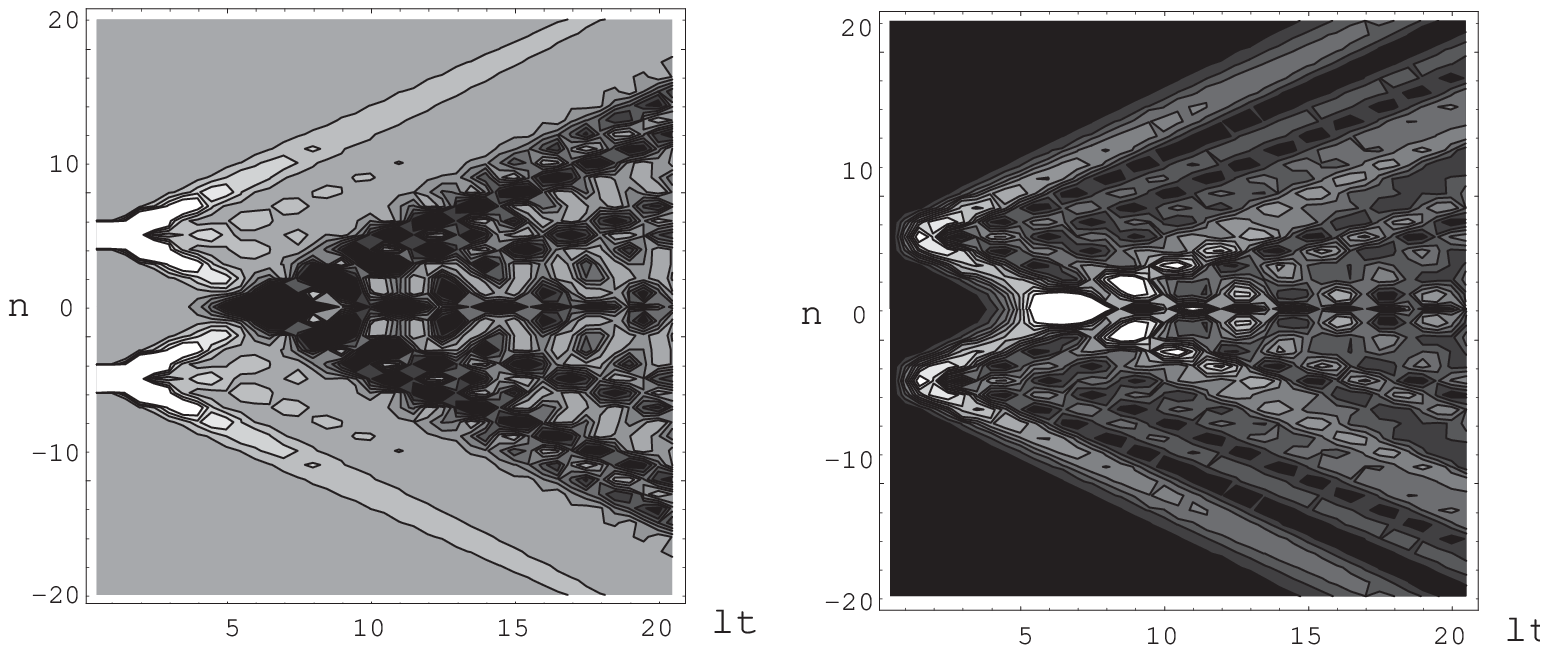}
\end{center}
\caption{{\em Upper panel}: Time evolution of the concurrence between sites $-n$ and
$n$ for the initial state $\ket{\Phi^\varphi_{-5,5}}$. The plot
is cut at $0.2$ in order to make the revival after the crossing
visible. {\em Lower panel}: Fidelity (white:high, black:low)
of the states $\ket{\Phi^{\varphi_{opt}}}$ (left)
and $\ket{\Psi^{\varphi_{opt}}}$ (right) in the density matrix
$\rho^{(2)}_{-n,n}(t)$ for the same initial condition as above.} \label{concPhi}
\end{figure}

\subsubsection{Propagation of states in the triplet sector}

We next analyzed the propagation of the state
\beq
\ket{\Phi_{i,j}^{\varphi}} =
\frac{1}{\sqrt{2}} ( \I + e^{i \varphi} c_i^{\dagger}
c_j^{\dagger} ) \, \ket{\Downarrow} \; . \label{statiPhi}
\eeq
These are not single-particle states and since they
are superpositions of components pertaining to different 
spin sectors, one cannot take full advantage
of the conservation of the magnetization. As a result, 
the concurrence is given by $C=\max\{0,C^{(1)},
C^{(2)}\}$;
a concurrence of the form $C^{(2)}$ indicates that
$\ket{\Psi}$-like correlation arises, while for $C^{(1)} > C^{(2)}$
the correlations between the two selected sites is more
$\ket{\Phi}$-like. This already suggests that changes in the propagated type of 
entanglement could occur.
\\
The entanglement still propagates with the sound velocity $\sim \lambda$
along the chain, as can be seen in Fig.~(\ref{concPhi}), where $C_{-n,n}$ is
displayed for an initial state $|\Phi^\varphi_{-5,5}\rangle$. 
At the intersection between the sites of the initial
Bell state a revival takes place, and
the entanglement then also spreads out from this intersection point.
In contrast to the previous case,
when the two initially entangled sites are separated by an 
odd number of spins, the propagating quantum correlations change character.
This is seen from the fidelity of the
Bell states $|\Phi^{\varphi'}_{m,n}\rangle$ and 
$|\Psi^{\varphi'}_{m,n}\rangle$ in the time evolution of the initial 
Bell state $|\Phi^\varphi_{i,j}\rangle$.
These two fidelities are maximized by a proper choice
of the phases\cite{AOXYdyn} 
which depend only on the initial and final positions $i$, $j$, $m$ and $n$
and are shown in the bottom of Fig. (\ref{concPhi}) --
left plot: $|\Phi^{\varphi_{opt}}\rangle$; right plot:
$|\Psi^{\varphi_{opt}}\rangle$ -- for $i=-j=5$ and $n=-m$.
Since the fidelity of $\ket{\Phi^\varphi}$ in the vacuum is $0.5$
due to its component $\ket{\downarrow \downarrow}$, it is seen 
that after the crossing point the state has
become $\ket{\Psi}$-like. In fact, at the single-site crossing, the
amplitude for two parallel spins cannot survive and the outgoing
states of this scattering event only contain antiparallel spins.
If the initially entangled spins are separated 
by an even number of sites, the crossing involves two sites and the
character of the state is preserved.

\subsection{$\gamma \ne 0$: Singlet onto the vacuum}
\label{section:generic}
We now study the model for generic $\gamma$ and infinite chain. 
In contrast to the previous section, 
here the complexity of the computation grows with the distance $d$ of
the sites because the dimension of the Pfaffian expression
for the correlation functions is $2d\times 2d$.
For the critical Ising model it turns out to be sufficient considering
$d=1$, since the concurrence vanishes for larger distances.
The initial state is the singlet  $|\Psi_{1,2}^\pi\rangle$, Eq.~(\ref{wavefunction}).

\subsubsection{Concurrence}

\begin{figure}
\begin{center}
\includegraphics[width=.26\linewidth,angle=-90]{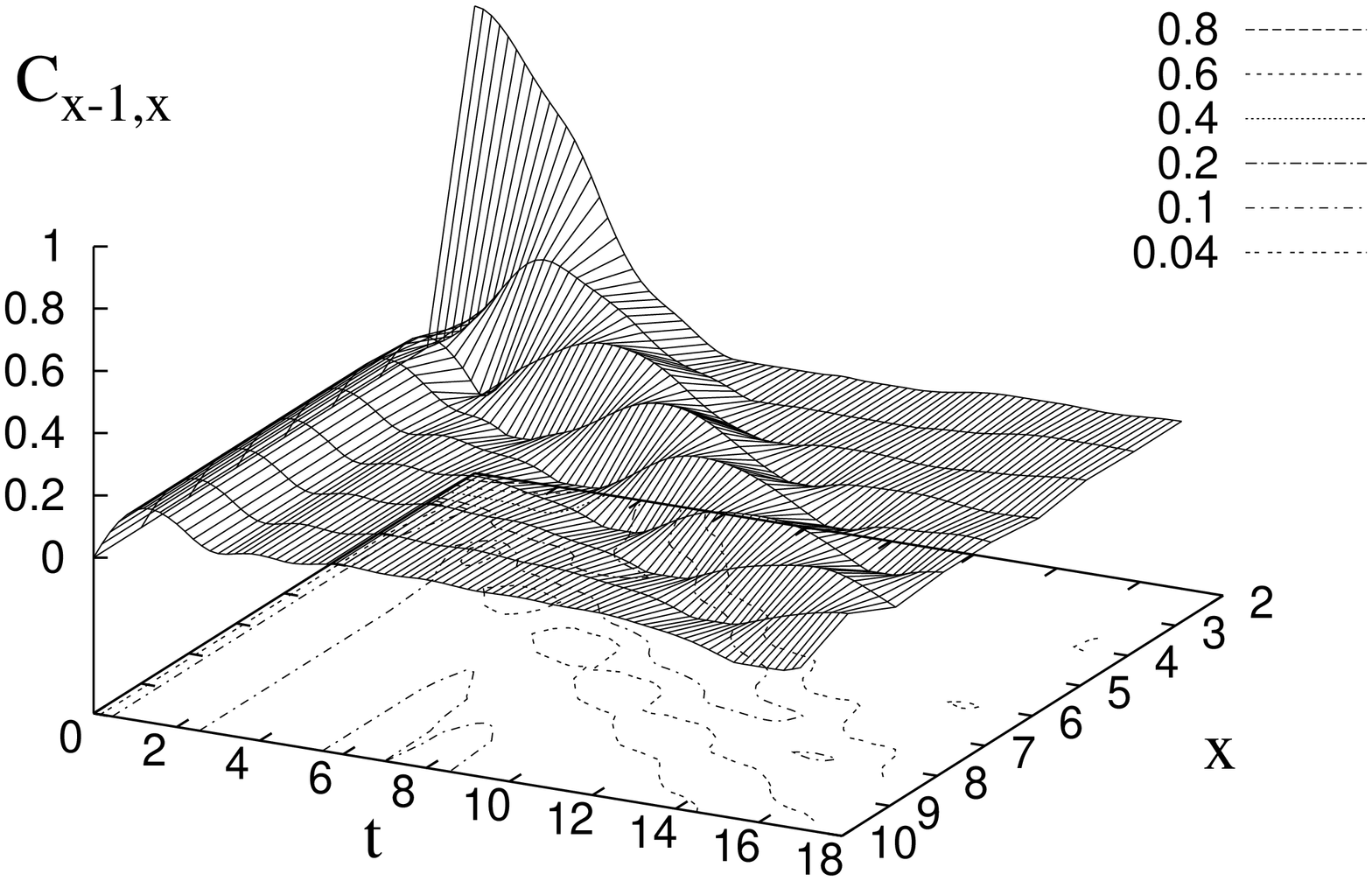}
\includegraphics[width=.26\linewidth,angle=-90]{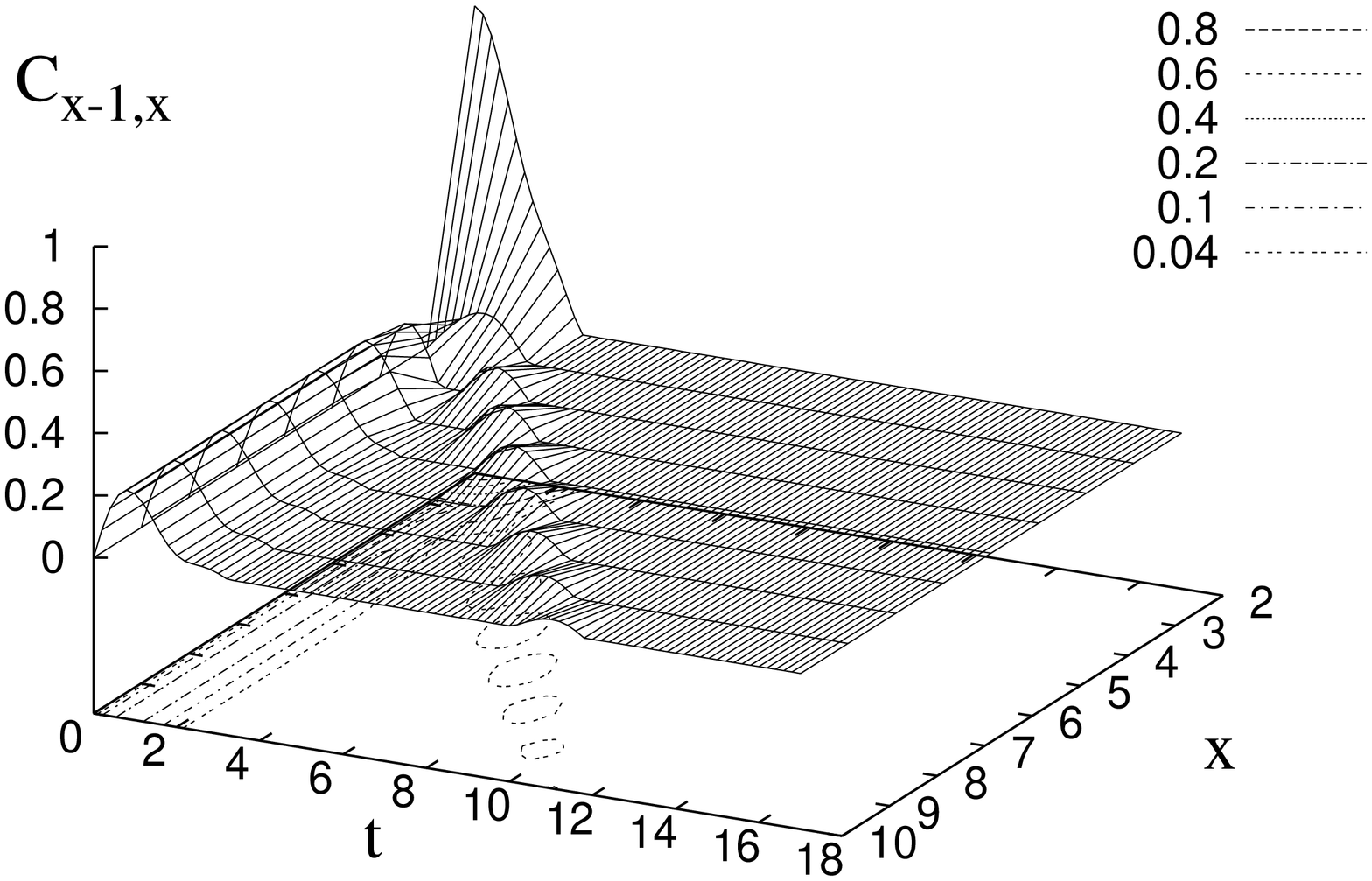}
\includegraphics[width=.26\linewidth,angle=-90]{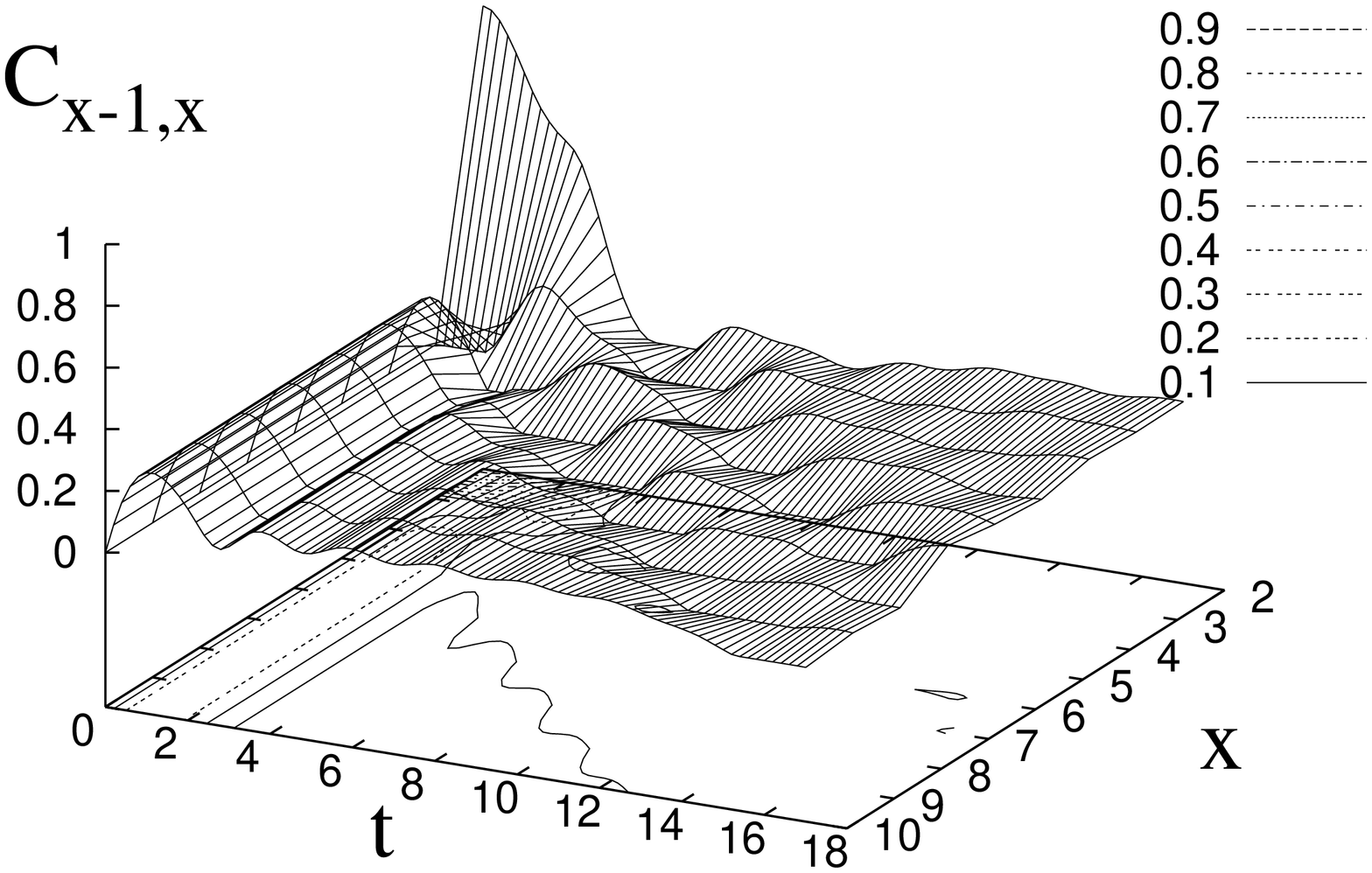}
\includegraphics[width=.26\linewidth,angle=-90]{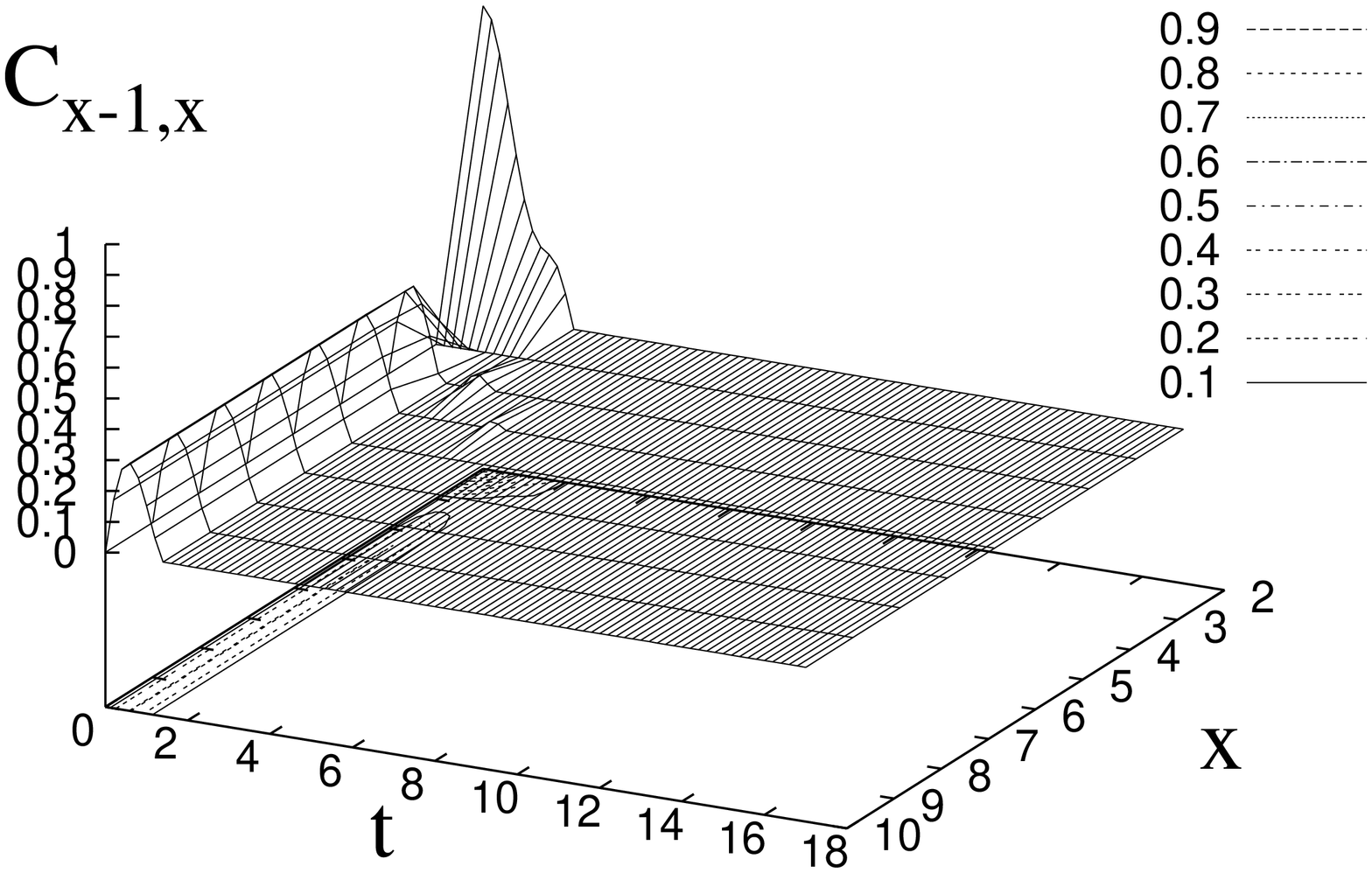}
\includegraphics[width=.26\linewidth,angle=-90]{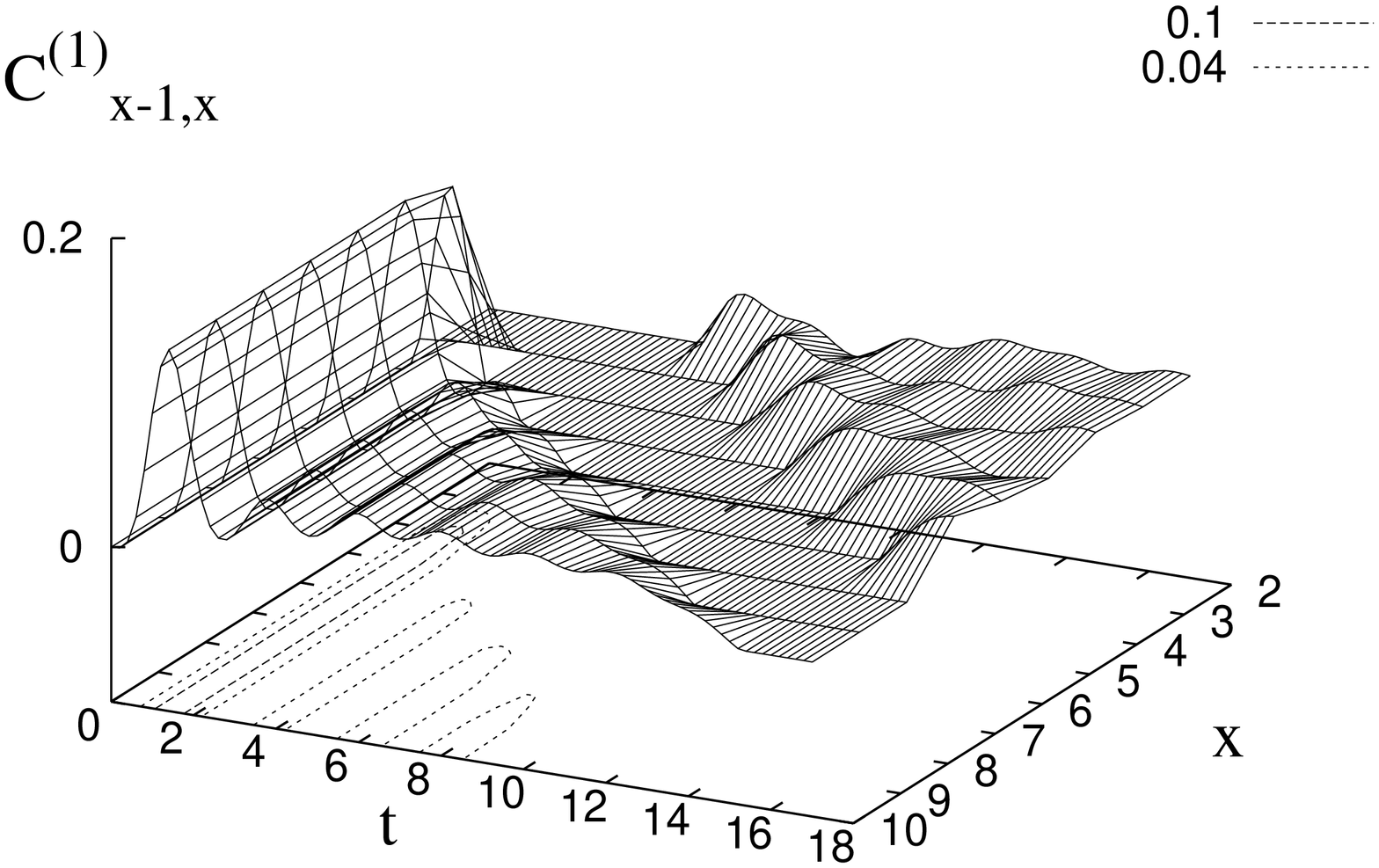}
\includegraphics[width=.26\linewidth,angle=-90]{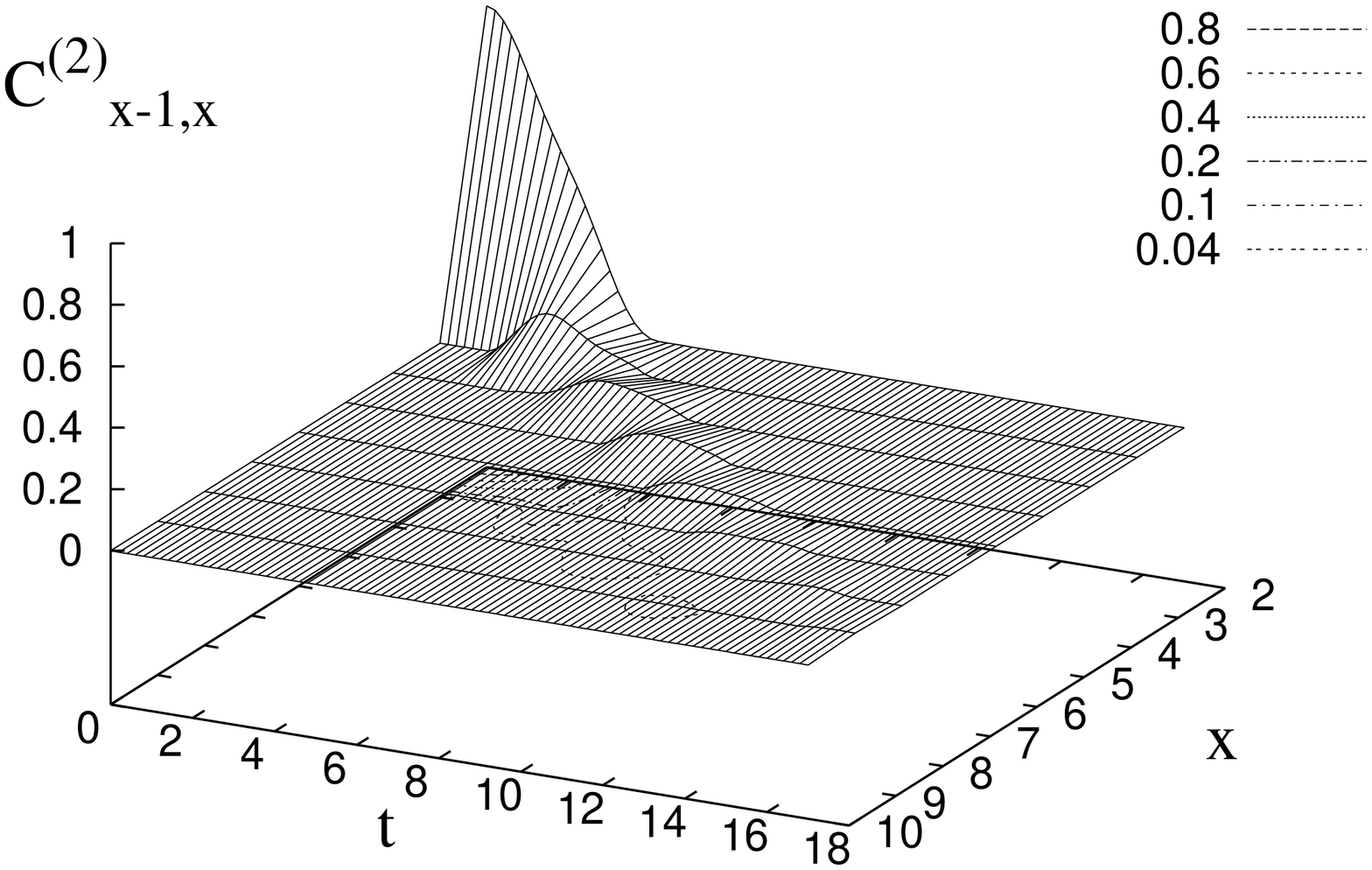}
\includegraphics[width=.26\linewidth,angle=-90]{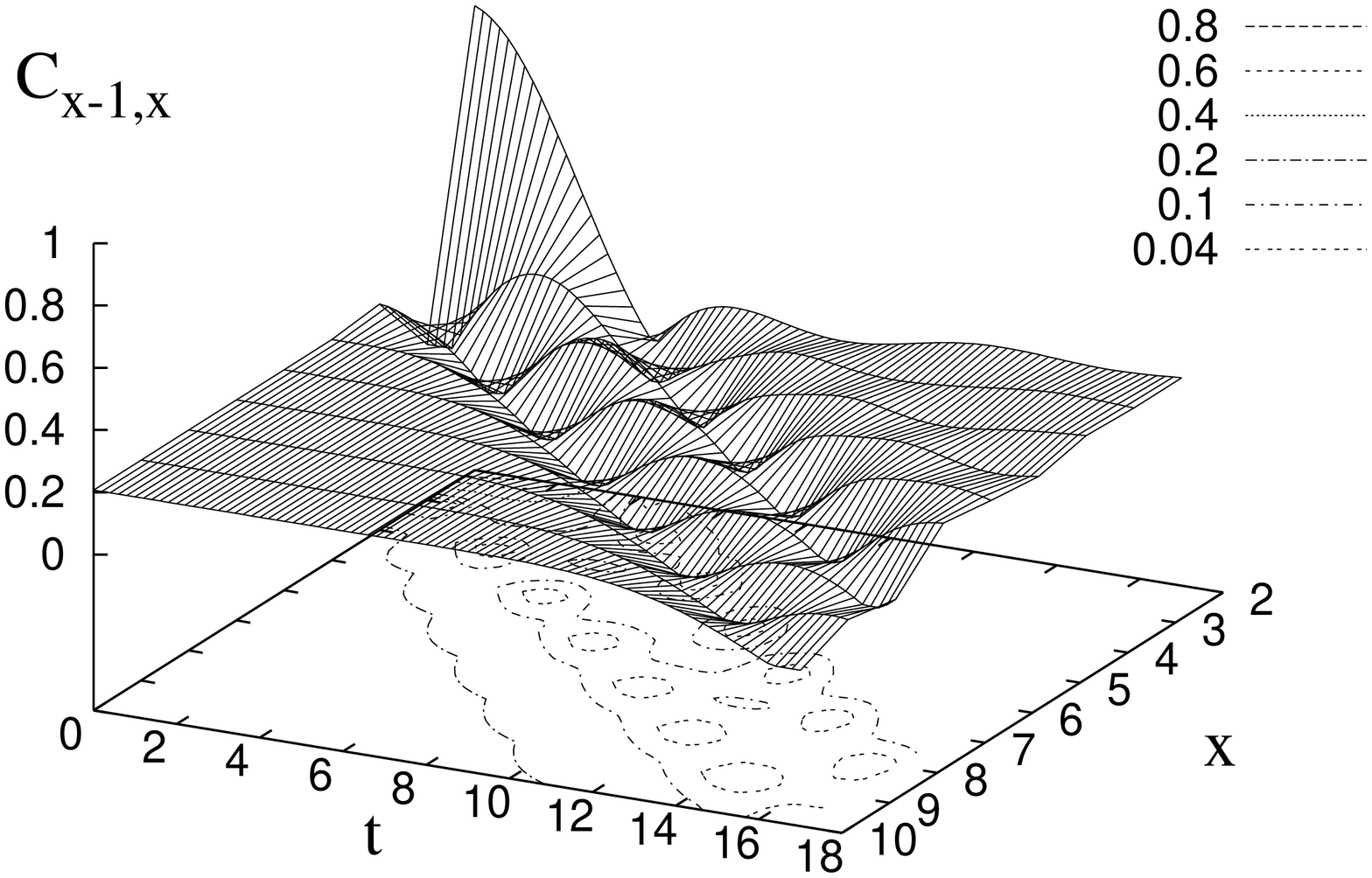}
\includegraphics[width=.26\linewidth,angle=-90]{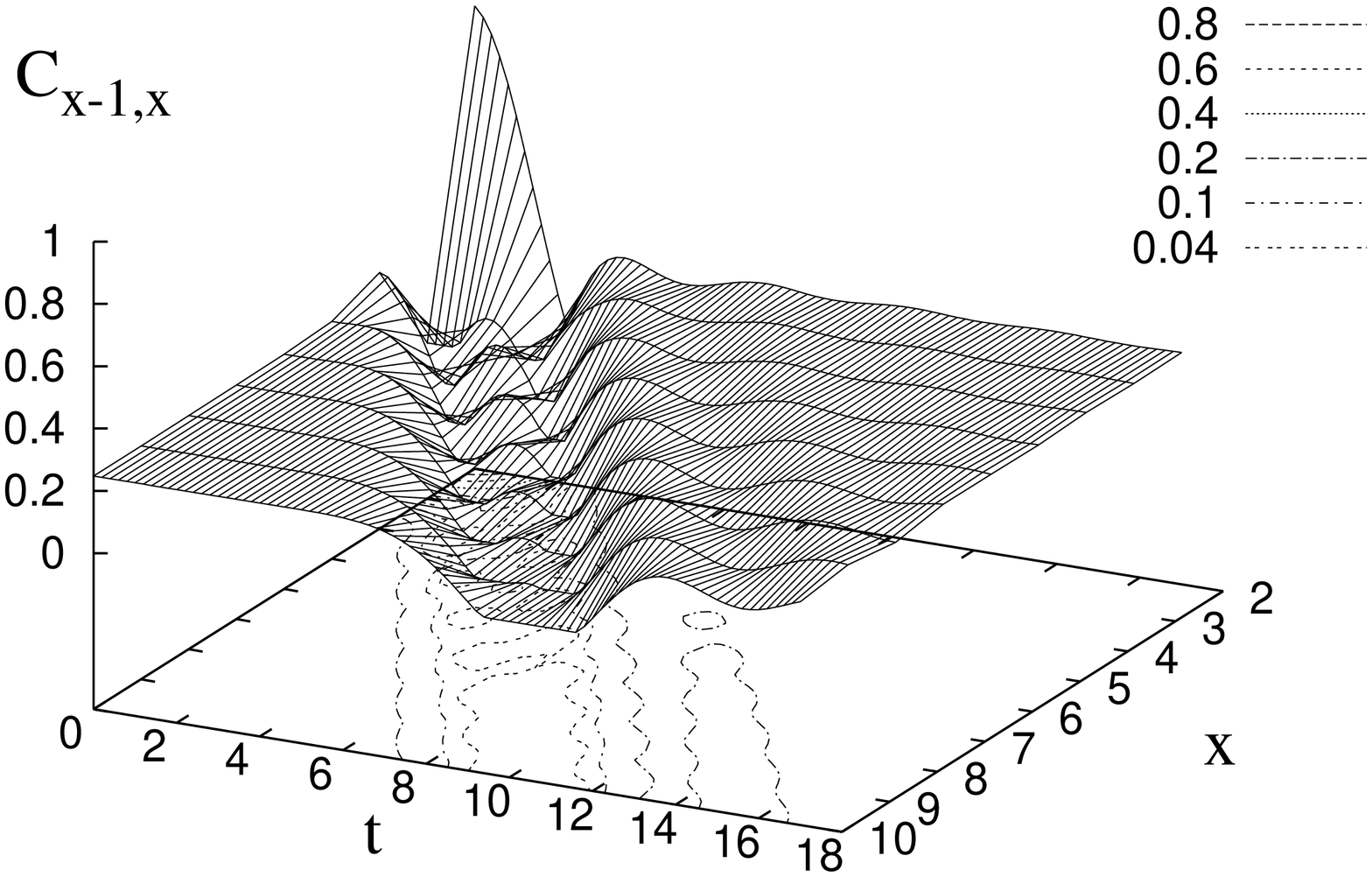}
\end{center}
\caption{
{\em Upper two panels}: The nearest neighbor concurrence for medium  $\lambda=0.5$ (left)
anisotropy $\gamma=0.5$
and $\lambda=0.5$ and $1$ (left and right, respectively).
The propagation is stronger damped than for small or zero anisotropy.
There is a creation from the vacuum which reaches the propagating pulse.
At the critical coupling (right), the nearest neighbor
concurrence from the initial singlet dies out immediately and so does
the vacuum creation. For the Ising model, only few bumps are residues of a propagation .
{\em 3'rd panel}: The square of $C^{(1)}$ (left) and $C^{(2)}$ (right) for the 
Ising model at $\lambda=0.5$ demonstrate that the propagating signal is in the 
singlet sector, whereas the background is in the triplet sector.
{\em Bottom panel}: 
$C_{x-1,x}$ for the perturbed ground state and the Ising model: $\lambda=0.5$ (left)
and $\lambda=0.9$ (right). There is a background concurrence 
corresponding to the ground state value (see Fig. \ref{separable-point}).
A propagating signal is seen in a valley of extinction.}
\label{C1-Vac}
\end{figure}

Figure \ref{C1-Vac} shows the nearest neighbor concurrence
for $\gamma=0.5,\ 1.0$ (1st and 2nd panel)
and $\lambda=0.5,\ 1.0$ (left and right, respectively). 
A rough estimate of the propagation velocity can be taken from 
the contour lines in the plot.
It coincides with the sound velocity, which is roughly $\lambda$
as for the isotropic model. For increasing $\gamma$ 
it slightly decreases for eventually returning to $\lambda$\cite{AOXYdyn,Stolze,Noeppert}. 
The concurrence on the
original singlet position decays quickly and the damping of oscillations is
stronger for increasing $\lambda$.
We notice an instantaneous signal which, 
sufficiently far away from the initial singlet position is 
spatially uniform. 
This phenomenon reflects the creation of entanglement from the vacuum,
which is characteristic for the anisotropic XY models. 
It is originated from the double spin
flip operators $\gamma\lambda/2\, \sum_i s_i^+ s_{i+1}^+$ 
(and its Hermitean conjugate), which are absent in the isotropic model.
The initial slope and the type of Bell state created agree well with simple
perturbative arguments.
This pure vacuum signal dies out very quickly. 
Towards the critical coupling and the Ising model, 
the damping of the concurrence propagation gets stronger. 
The vacuum signal survives much longer for
medium $\lambda$ and $\gamma\rightarrow 1$ such 
that for $\gamma=0.5$ and $\gamma=1$ it interferes with the 
propagating singlet. 
Although the damping of the propagation becomes 
stronger at the critical coupling,
nevertheless it is of pure dynamic origin 
and not related to the quantum phase transition, since
the initial state is not
the ground state (where the critical behaviour is encoded in). 
Consistently, the damping turned out
to be independent of the size of the chain.
\\
The concurrence for $\gamma=0.5$ is shown in the top pannel of 
Fig. \ref{C1-Vac}:
we see a clear propagation of the concurrence, which is only
slightly stronger damped than for the isotropic model, but there is a
creation of entanglement from the vacuum.
A ``shoulder'' appears in the singlet peak of
$C_{i,i+1}$, i.e. on the original singlet position, which is due
to triplet-type entanglement\cite{AOXYdyn}.
The squares of the preconcurrences $C^{(1)}$ and $C^{(2)}$, 
shown in the 3rd panel of Fig.~\ref{C1-Vac},
alike the analysis of the fidelity of the Bell states in the propagating signal~\cite{AOXYdyn}, 
demonstrate that the propagation is of the same type as the initially created Bell state
(here in the singlet sector), whereas the concurrence created from the vacuum
is of triplet type.
\begin{figure}
\begin{center}
\includegraphics[width=.3\linewidth,angle=-90]{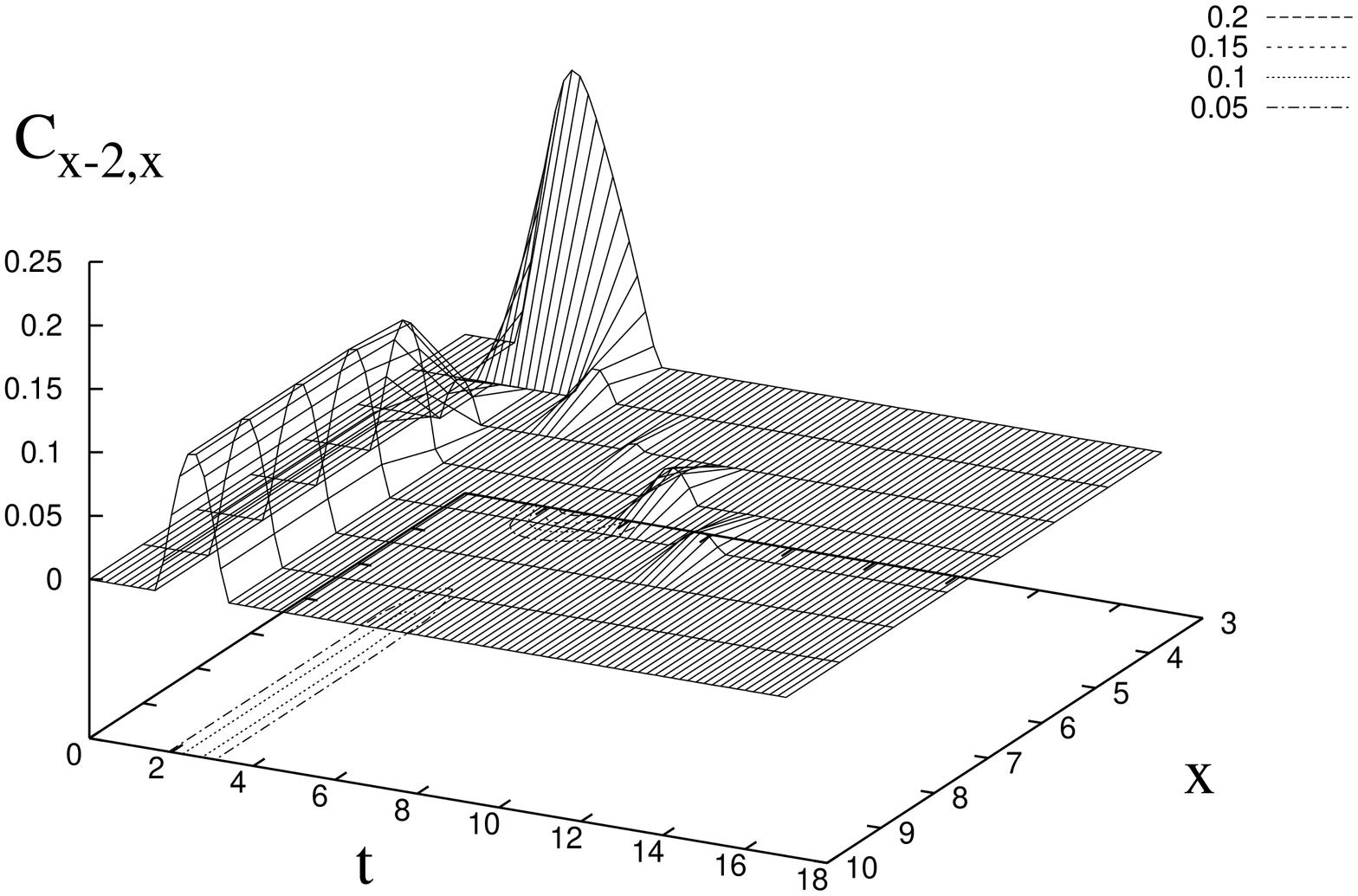}
\includegraphics[width=.3\linewidth,angle=-90]{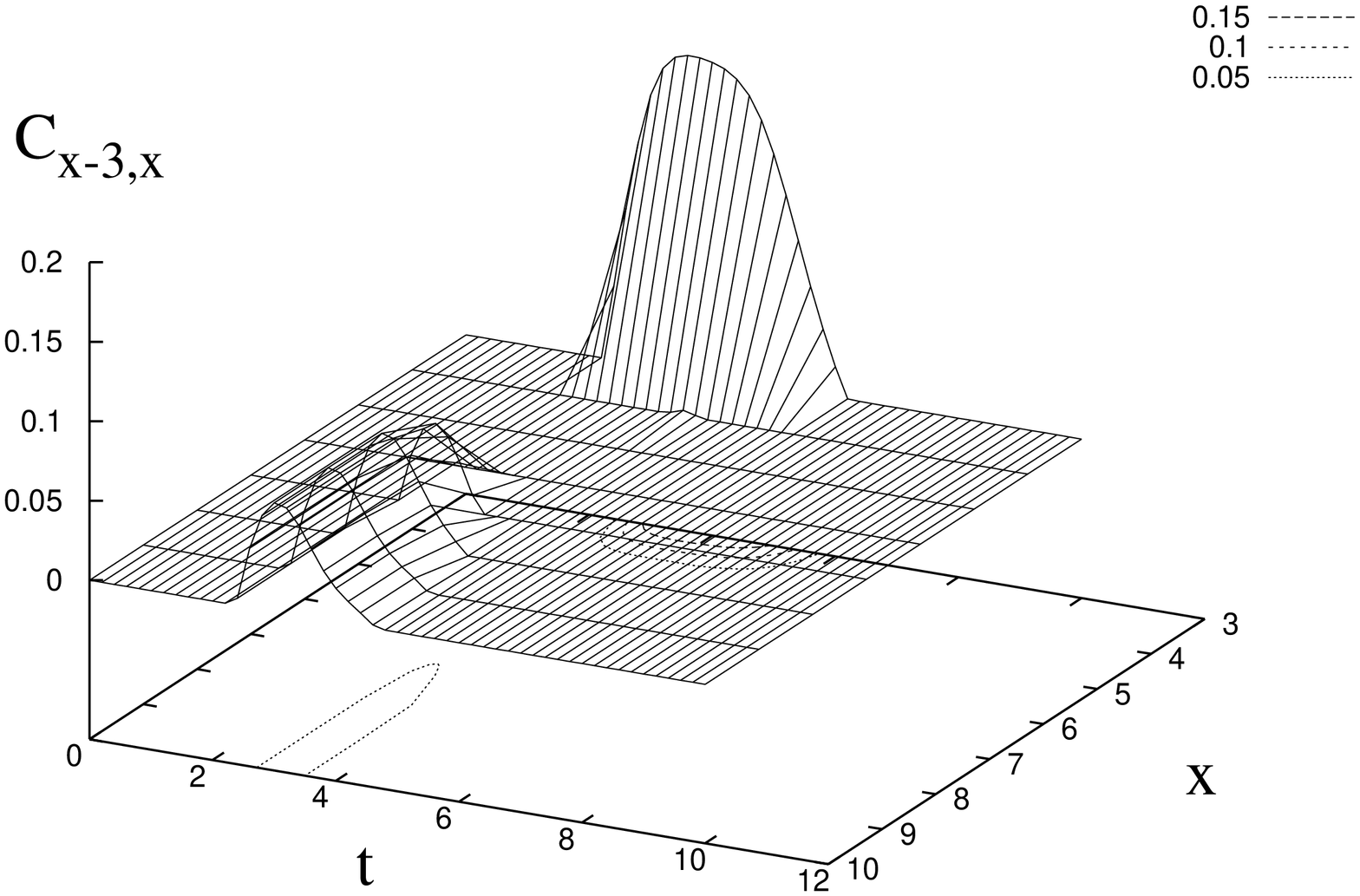}
\end{center}
\caption{
{\em Left -} The next nearest neighbor concurrence $C_2$ for the Ising model
far from the critical coupling. It is zero at and close to the critical coupling.
{\em Right -} The next-next nearest neighbor concurrence $C_3$ for the Ising model
far from the critical coupling. Also $C_3$ vanishes at and close to the critical coupling.
At $\lambda=0.5$ we see a considerably large signal, which is due to
an EPR-type propagation of a ``split'' singlet.}
\label{C2andC3}
\end{figure}
For the transverse Ising model (2nd panel in Fig. \ref{C1-Vac}), 
the shoulder on the original singlet position
and the vacuum creation gets even more pronounced, but all the signals 
die out much quicker. 
For $\lambda=0.5$ one cannot speak any more of a clearly propagating
entanglement signal.
At the critical coupling, the propagating signal almost completely disappeared.
The next-nearest neighbor 
and next-next-nearest neighbor concurrence are shown in Fig. \ref{C2andC3}.
Both show a narrow wall created from the vacuum, which for $C_3:=C_{x-3,x}$ 
is broader. $C_3$ unveils an additional feature: it shows a large contribution
at $t$ around $4$ at $x=3$. 
It indicates an EPR-type propagation, which
we observed already for the isotropic model (see Fig. \ref{treconc}).
Considering the velocity $\lambda$, the
time of appearance indicates that the two fragments have to cross first.
At and near the critical coupling $C_2$ and $C_3$ identically vanish on the
domain of the demonstrated plots.

\subsubsection{The global tangle}

Is it possible to understand how the entanglement, originally stored
into the singlet, is going to share among all the spins in the chain? 
In order to get an idea about what might happen, one can study the one-tangle. 
\begin{figure}
\begin{center}
\includegraphics[width=.23\linewidth,angle=-90]{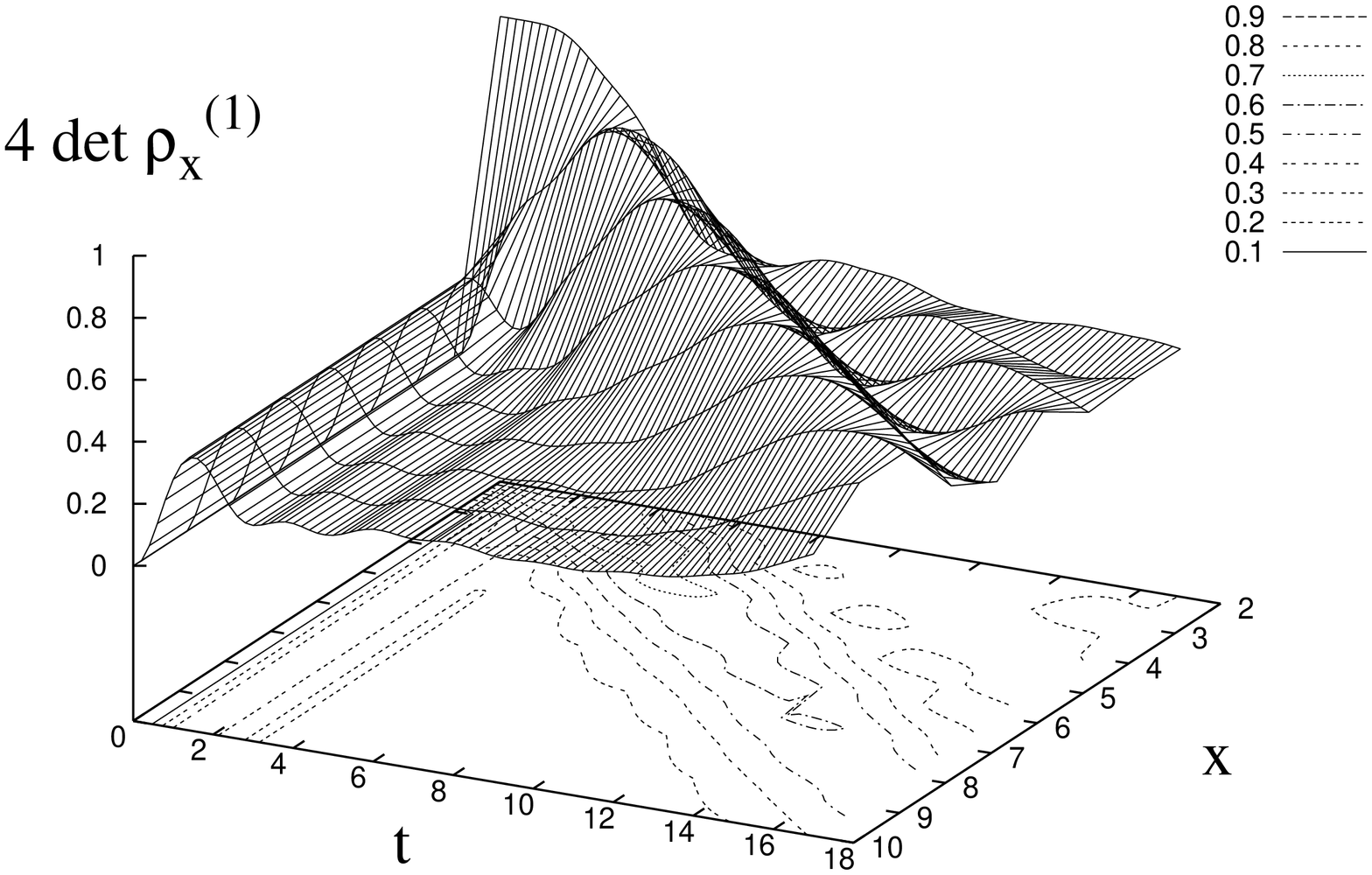}
\includegraphics[width=.23\linewidth,angle=-90]{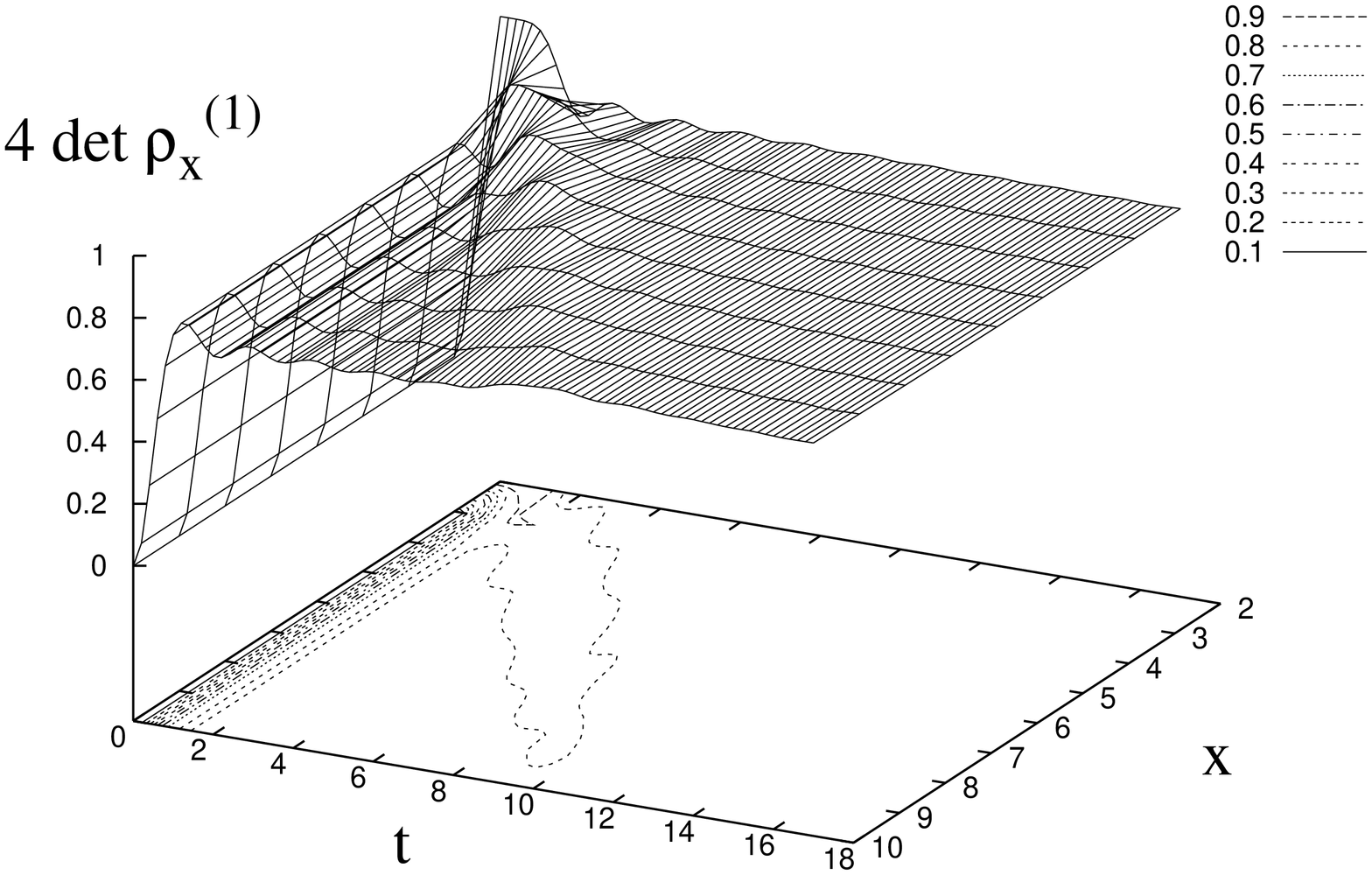}
\includegraphics[width=.23\linewidth,angle=-90]{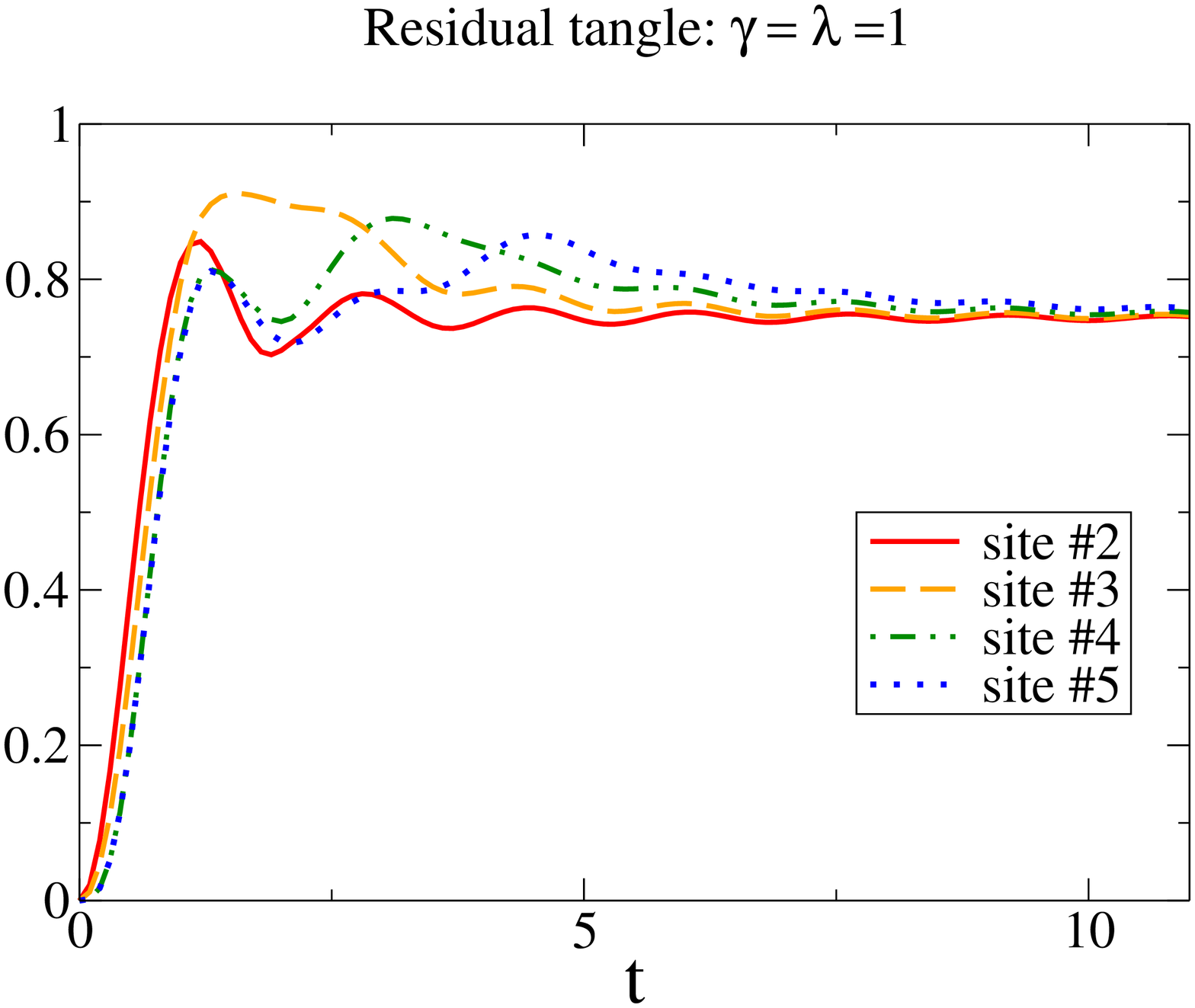}
\end{center}
\caption{The total tangle of site $x$ for $\gamma=1$. 
In contrast to the nearest neighbor concurrence there is a 
clear propagating signal here fro $\lambda=0.5$ (left). It is mounted 
on top of a non-zero background signal coming from the vacuum. 
At critical coupling (middle) the vacuum background of the one-tangle grows up to about $0.75$; the propagation is hardly visible. {\em Right}: The CKW conjecture applies: the ``residual tangle''
indeed is positive.}
\label{totaltangle}
\end{figure}
For small anisotropy parameter, this quantity is qualitatively 
very similar to the concurrence: there is an entanglement 
wave, propagating with velocity $\lambda$.
For growing $\gamma$, the wave is suppressed in favor of a homogeneous
growth of the one-tangle which saturates
on very short timescales.
This signal cannot be due to the initial singlet, of course, but it 
is created from the vacuum.
This was confirmed by looking at the total entanglement created solely from 
the vacuum, i.e. the initial state being the vacuum.~\cite{AOXYdyn}
The right-most plot in Fig.\ref{totaltangle} shows the residual tangle
for the Ising model at $\lambda=1$; there is no violation of the CKW conjecture.

\begin{figure}
\begin{center}
\includegraphics[width=.23\linewidth,angle=-90]{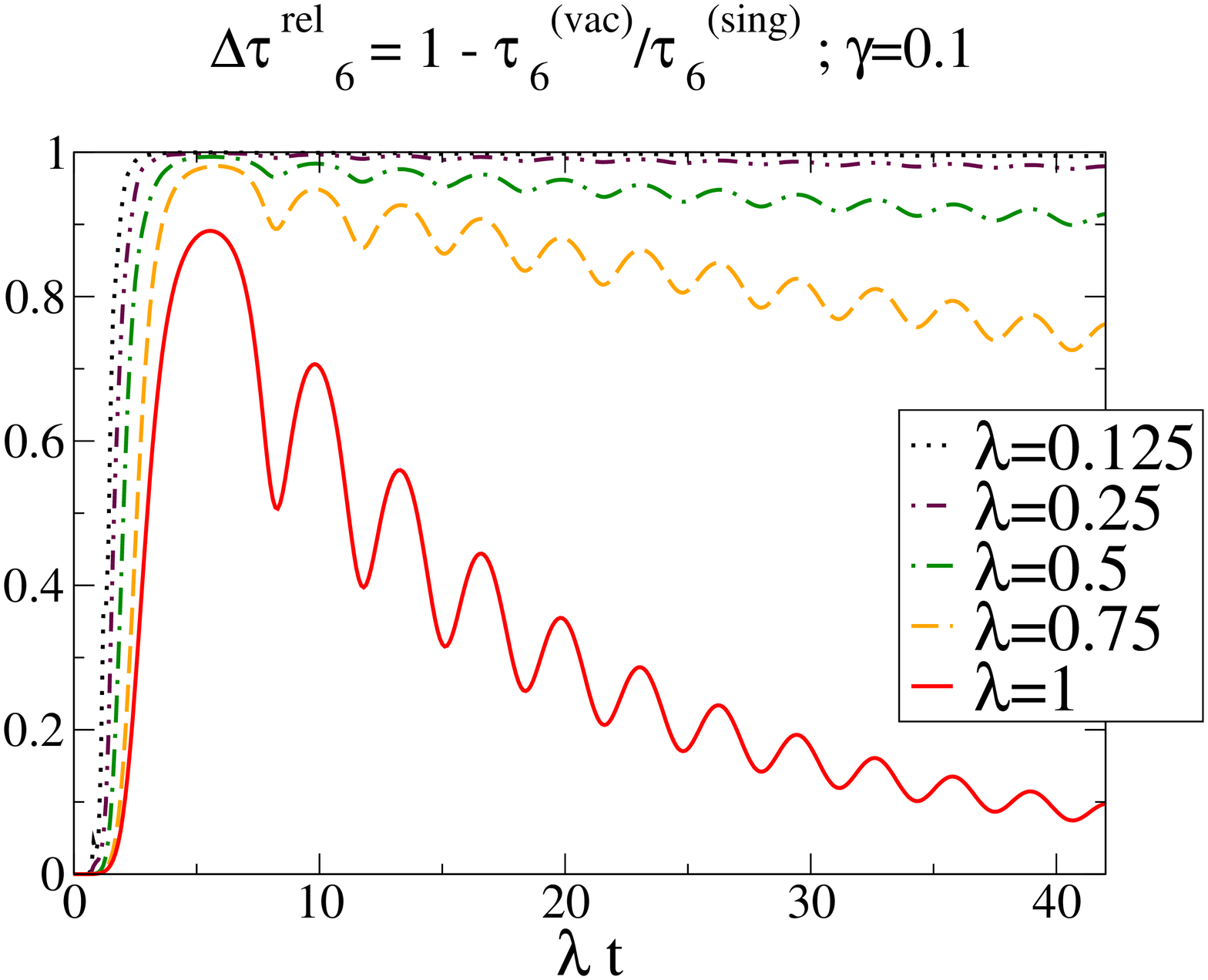}
\includegraphics[width=.23\linewidth,angle=-90]{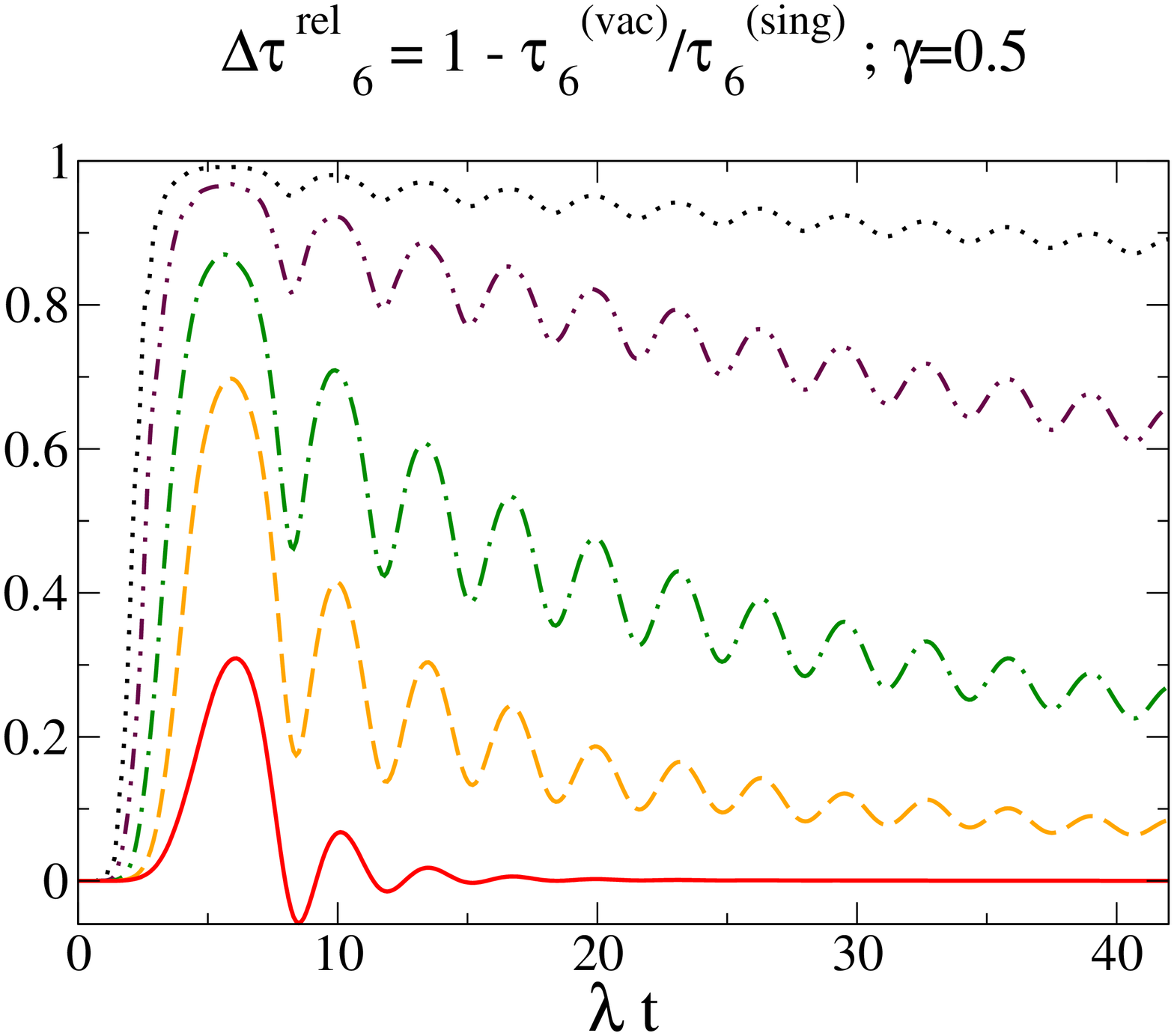}
\includegraphics[width=.23\linewidth,angle=-90]{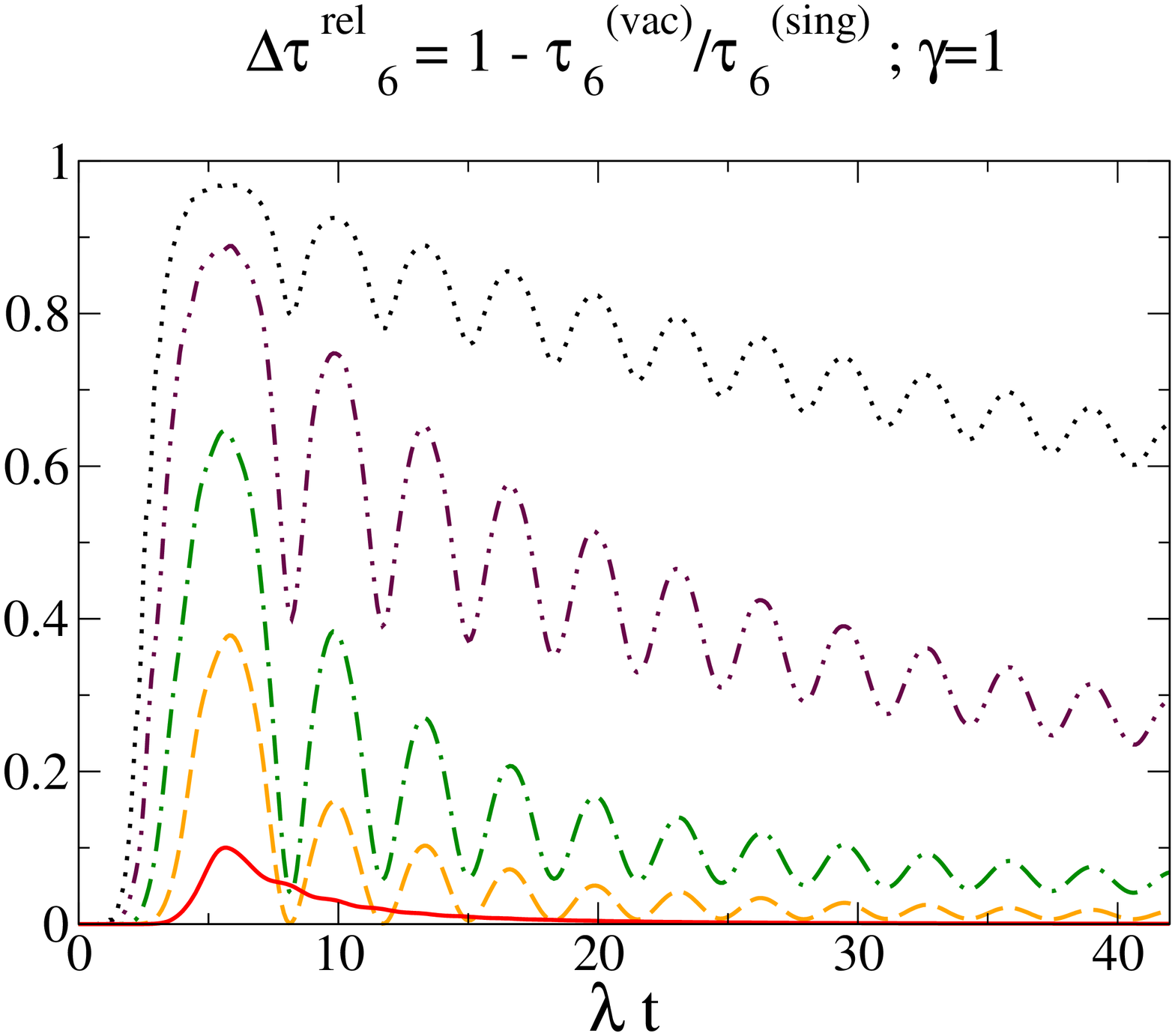}
\includegraphics[width=.23\linewidth,angle=-90]{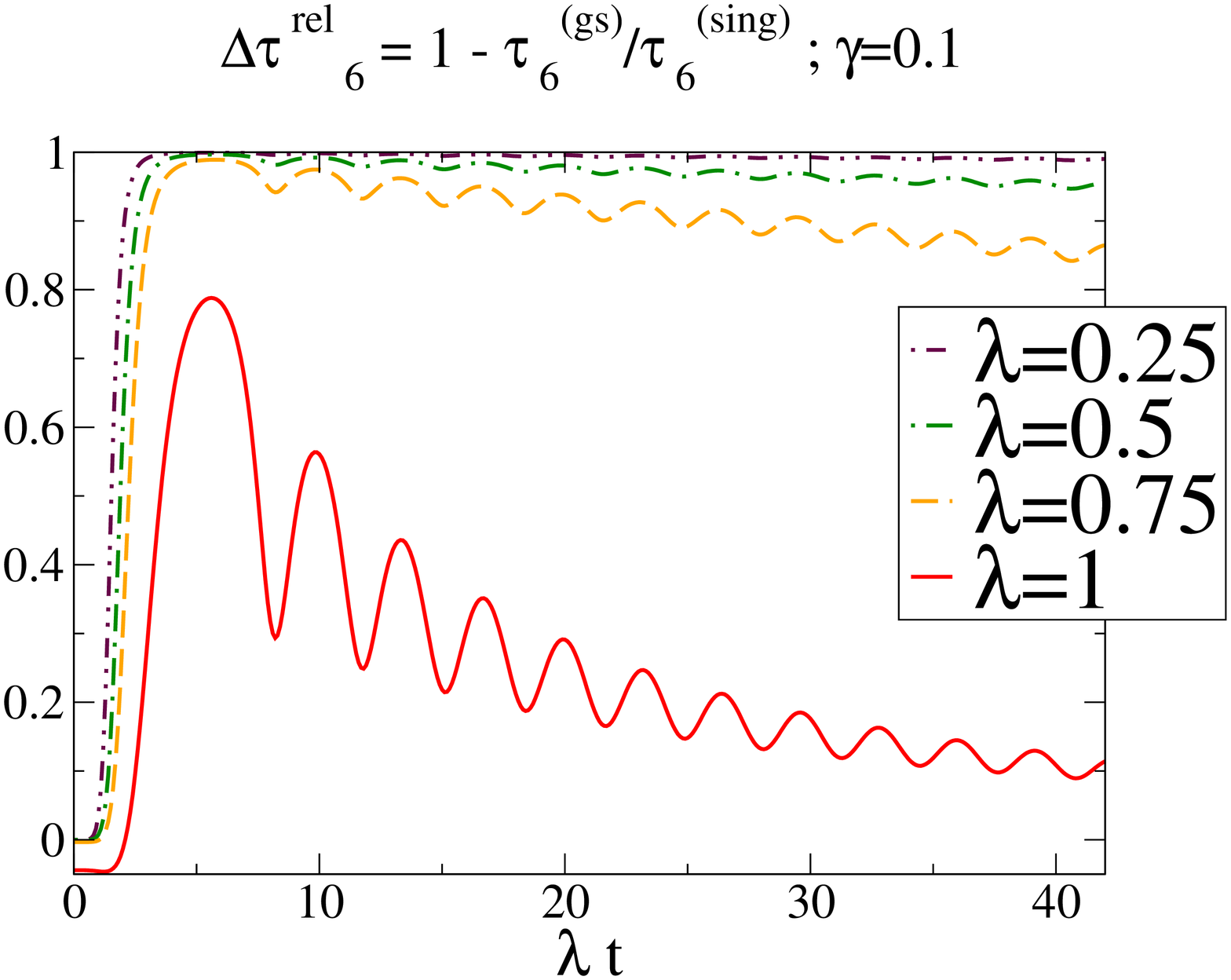}
\includegraphics[width=.23\linewidth,angle=-90]{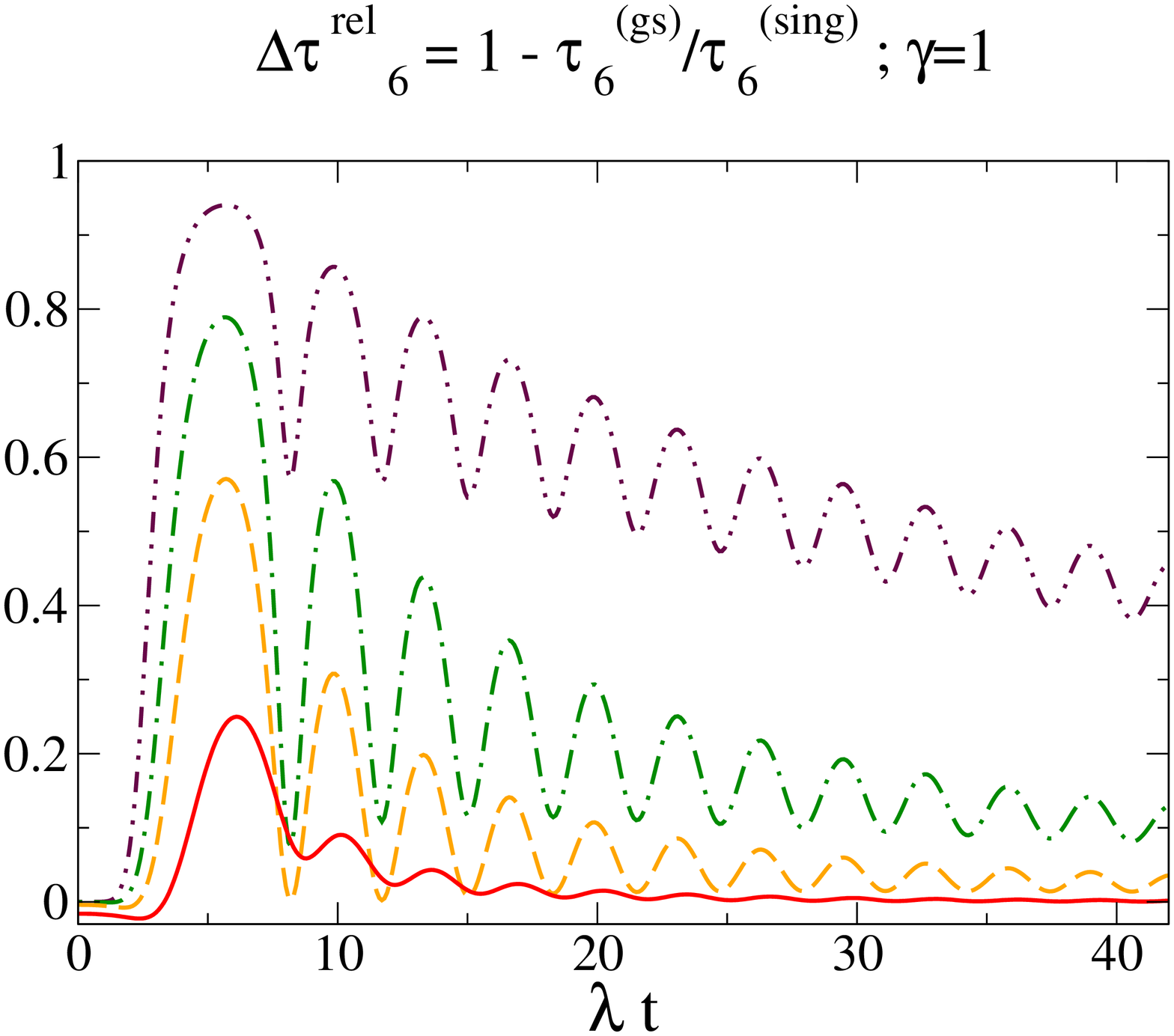}
\includegraphics[width=.23\linewidth,angle=-90]{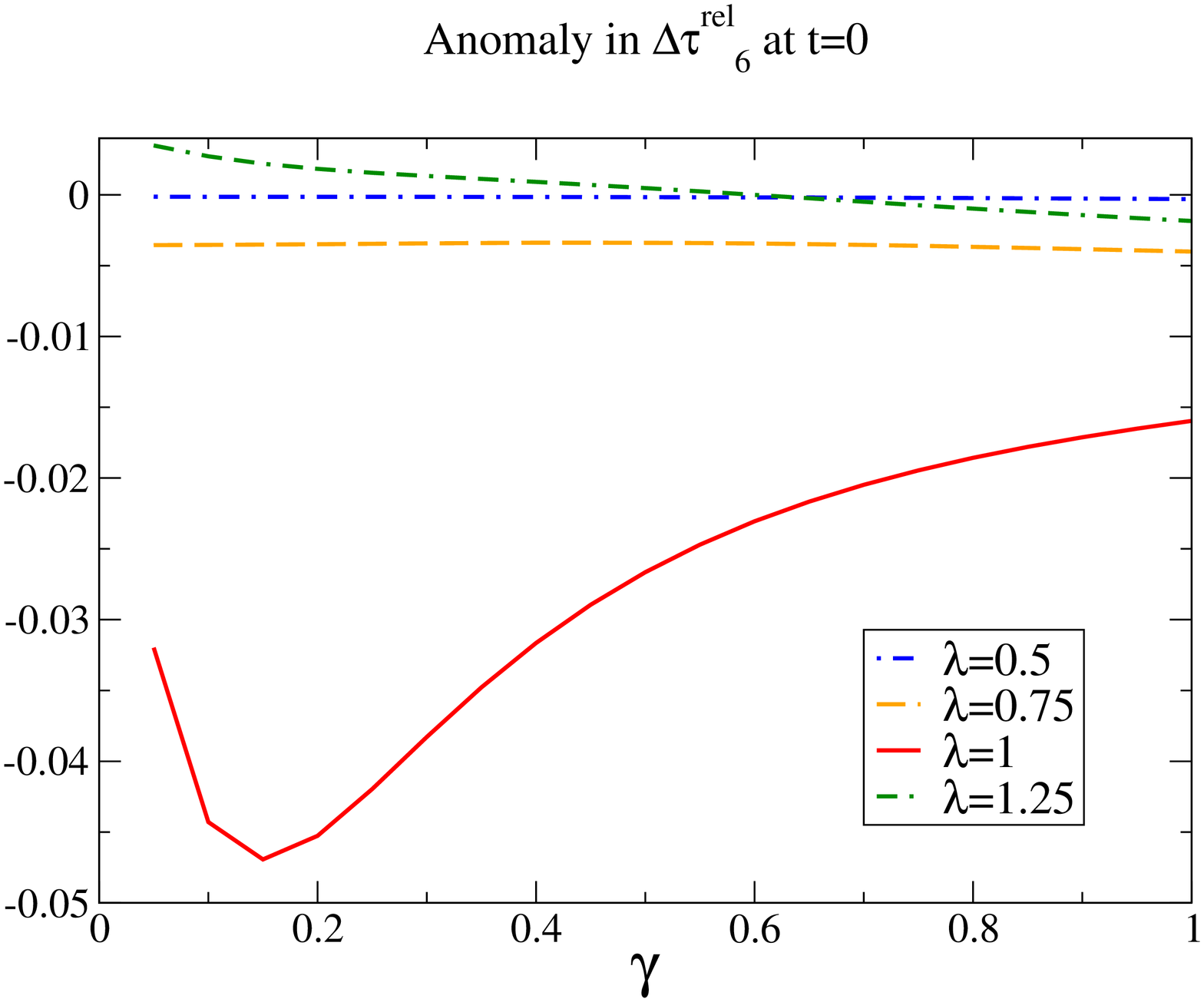}
\end{center}
\caption{{\em Upper panel}: The relative tangle deviation $\Delta\tau^{rel}_6$ from the Vacuum 
tangle at site number $6$ for different values of $\lambda$ for a fixed 
value of $\gamma$: $\gamma=0.1$ (left), $\gamma=0.5$, and $\gamma=1$ (right). 
It is plotted as a function of the reduced time $\tau=t/\lambda$;
it is nicely seen that the oscillation frequency grows linearly with 
$\lambda$. {\em Lower panel}: The same as above but for the ground state and 
$\gamma=0.1$ (left) and $\gamma=1$ (center).
The oscillation frequency grows linearly with $\lambda$. Right: anomaly at t=0: it demonstrates the non-local impact of the singlet-type perturbation at the critical coupling.}
\label{reldeltangle-Vac-of-l}
\end{figure}

Next, we will study the relative deviations in the one-tangle respect to the 
time-evolved vacuum
\beq\label{deltangle}
\Delta \tau^{rel}_j:=1 - \bigfrac{\det \rho_{(vac)}^{(1)}}{\det \rho_j^{(1)}}\; .
\eeq  
We want to stress that the time 
evolution operator does not in general preserve the relative order induced 
by an entanglement measure on the Hilbert space. This manifests
in negative values of $\Delta \tau_j$.
Eventually, this is due to the 
fact that superposing (as well as mixing) orthogonal maximally 
entangled states of the same type diminishes the entanglement.
This may lead to a negative $\Delta \tau^{rel}$.
It is worth noticing that the singlet state is inserted into the vacuum;
this is not a superposition of the vacuum and some other state.
We choose the same parameter range as above.
Along the axis $t=0$ and $\lambda=0$ and for site numbers larger than
two (and smaller than one) we have that $\det \rho_j^{(1)}=0$.
In these cases, also $\det \rho_{(vac)}^{(1)}=0$, and we chose
the plotted value being zero in these cases.
The analysis of $\Delta \tau^{rel}_j$ tells us that for small anisotropy
and sufficiently far from the critical coupling 
the global tangle is dominated by the local perturbation of the
vacuum by the singlet (Fig.~\ref{reldeltangle-Vac-of-l}: top),
meaning that the total tangle is concurrence
dominated. For the isotropic $XY$ model, the global tangle 
was given entirely by the sum of the $2$-tangles such that the 
CKW-conjecture would conclude that there is no higher tangle 
contained in the system.
In the presence of a small anisotropy, this is no longer true,
in particular near to the critical coupling $\lambda=1$
(Fig.~\ref{reldeltangle-Vac-of-l}).
Rising anisotropy and $\lambda\rightarrow 1$ enhance the vacuum domination.

\subsection{$\gamma \ne 0$: Singlet-type perturbation of the groundstate}
\label{section:gammaground}

One could wonder, whether or not the
results in the preceding sections were specific to the vacuum state
or what would happen for different states.
We therefore discuss in this section the propagation of a singlet-like
perturbation of the groundstate $\ket{GS}$, i.e. the
time evolution of the initial state
\beq\label{initialstate:GS}
\ket{S}_g:=\bigfrac{1}{\sqrt{2}}(c_1 - c_2)\ket{GS}.
\eeq
We note that this state  
differs from a singlet on the ground state since the operators $c_i$ 
create global excitations. 
It turned out that the resulting state is nonetheless
very similar to the ground state itself~\cite{AOXYdyn}.
For the isotropic model $\ket{GS}\equiv\ket{\Uparrow}$ and hence
the dynamics is the same as on the vacuum.

Apart from the propagation of the pulse
with velocity about $\lambda$, there are several qualitative differences
for the concurrence. 
One of them is that the propagating pulse and the initial Bell state 
on sites $1$ and $2$ is the triplet with zero magnetization. This
comes from subtleties in the Jordan-Wigner transformation~\cite{AOXYdyn}.
The concurrence plots in the bottom panel of Fig. \ref{C1-Vac}
show that the inital state is almost fully entangled on sites $1$ and $2$. 
This demonstrates how close is the ground state to $\ket{\Uparrow}$.
Two new features appear in the concurrence signal.
Firstly, the entanglement pulse propagates on top of a nonzero 
background level,
which in very good agreement coincides with the nearest neighbor 
concurrence of the ground state, being around $0.2$ at 
both, $\gamma=0.9$ and $\gamma=0.5$\cite{Osterloh02} 
(see Fig. \ref{separable-point}). This shows that not only is
the ground state very close to the vacuum, but that 
also the state $\ket{S}_g$ is very similar to the ground state
as far as nearest neighbor concurrence is considered.
Secondly, in contrast to the singlet on the vacuum, we here have to deal with
a propagation that eliminates the background concurrence. 
The latter feature gets more pronounced when approaching the Ising model 
and critical coupling (see bottom panel of Fig.~\ref{C1-Vac}).
This elimination is understood from the fact that joining entanglement
of the same type (here: two-site entanglement) in form of states that 
are orthogonal to those forming the entanglement already present, diminishes
the entanglement.
Whereas here the impact of the propagating concurrence pulse 
coming from the initial $0$-triplet is the stronger the closer we get to
the critical coupling and the quantum Ising model (the opposite of what we
observed for the initial singlet onto the vacuum), the situation
would be the same as far as the total signal along the propagation line
is concerned. This in fact gets more suppressed with growing 
$\lambda$ and $\gamma$. 
\\
For the global tangle we find essentially the same behavior
as for the singlet on the vacuum. 
Only at very short times the 
singlet-type perturbation deminishes the global tangle.
In the bottom panel of Fig~\ref{reldeltangle-Vac-of-l} we choose two different 
anisotropies $\gamma=$ $0.1$ and $1$ 
and compare $\Delta\tau^{rel}_j$ for different couplings $\lambda$.
The short-time behavior of $\Delta\tau^{rel}_6$
shows a marked anomaly at the critical coupling. This is a cursor of the non-local
impact of the singlet-type perturbation of the ground state, which is pronounced at
$\lambda_c=1$. At $\gamma=0$ all curves should eventually tend to zero.

\section{Conclusions}

In the present work we studied the effect of a singlet-type
perturbation on the entanglement of an infinite spin system. We
considered quantum $XY$ models for general anisotropy.
The dynamics of entanglement was studied as 
function of the distance to the local perturbation at
$t=0$, and of the reduced interaction strength $\lambda$ (up to the
common quantum critical point of the models at $\lambda=1$).
For this class of models we analized a conjecture formulated
by Coffman, Kundu, and Wootters quantifying the weight of the 
pairwise relatively to the global entanglement, measured by
$4 \det \rho_1 $.
In all cases the main propagating signal is in the same sector as that 
one initially created and the propagation velocity is in good agreement 
with the sound velocity of the model, which is roughly $\lambda$.
The isotropic model, i.e. anisotropy $\gamma=0$,
can be mapped onto a tight binding model, and consequently
entanglement propagates only, remaining pairwise. In addition we have an 
EPR-type propagation of the concurrence.
We found that states in the triplet sector (with concurrence $C=C^{(1)}$) 
are mutated after a crossing on a single site.
The global tangle and the concurrence (whose square is the $2$-tangle) 
satisfy the Coffman-Kundu-Wootters conjecture with zero residual tangle. 
This means that {\it the system contained only pairwise 
entanglement, measured by the concurrence}. 
The Hamiltonian does not create any entanglement; it distributes the
initially created pairwise entanglement.
Also for general anisotropy we found evidence for an EPR-type propagation. 
The propagation is suppressed, compared to the isotropic model.
The suppression is the stronger the closer the system is to
the critical coupling and the quantum Ising model.
For the latter we found a very rapid damping
of the singlet in the nearest neighbor concurrence.
For all larger distances we considered (up to $7$ lattice spacings) 
the concurrence is zero.
A peculiarity of the anisotropic models is the instantaneous
creation of concurrence from the vacuum all over the chain. 
It decays very quickly when approaching the quantum
critical coupling and getting close to the Ising model. 
Neither effect is of critical origin, though.
Comparing quantitatively  $4 \det \rho_1 $  and concurrence we 
conclude that {\it  for the anisotropic model  
the propagation of the concurrence is a small effect respect to 
the creation of higher tangles}. 
Medium interaction strengths
and/or small anisotropy favor the propagation of the singlet.
For the singlet-type perturbation on the ground state, 
the concurrence signal occurs along a ``valley'' in
the constant background concurrence.
The propagating extinction gets more enhanced with growing
anisotropy and approaching the critical coupling; 
for the critical Ising model a weak propagating signal in the valey remains.
The background nearest neighbor concurrence coincides with 
that of the ground state, indicating that the nearest neighbor
concurrence away from the initial perturbation is unaffected,
and that the Hamiltonian cannot notably create nearest neighbor concurrence
beyond that level. 
The dynamics of the total tangle is basically unchanged. 
In the short-time behavior of $\Delta\tau^{rel}_x$
an anomaly at $\lambda=1$
unveils the non-locality of the singlet-type perturbation of the ground state.
From the perspective of quantum information, our results can be 
read as a transfer of a unit of pairwise entanglement (an {\em e-bit})
within a spin chain by Hamiltonian action.
The so transported concurrence can then be distilled
at the destination point~\cite{Bose01,Horodecki97}.

\acknowledgments
The authors would like to thank G. Falci and J. Siewert
for helpful discussions. This work was supported by the EU (IST-SQUBIT, RTNNANO), 
RTN2-2001-00440, and HPRN-CT-2000-00144.

\end{document}